\documentclass[%
 aip,
 amsmath,amssymb,
 reprint,%
]{revtex4-1}

\usepackage{graphicx}
\usepackage{dcolumn}
\usepackage{bm}

\usepackage[utf8]{inputenc}
\usepackage[T1]{fontenc}
\usepackage{mathptmx}
\usepackage{etoolbox}

\usepackage[percent]{overpic}

\makeatletter
\def\@email#1#2{%
 \endgroup
 \patchcmd{\titleblock@produce}
  {\frontmatter@RRAPformat}
  {\frontmatter@RRAPformat{\produce@RRAP{*#1\href{mailto:#2}{#2}}}\frontmatter@RRAPformat}
  {}{}
}%
\makeatother
\begin{document}


\title[Majorana bound states with chiral magnetic textures]{Majorana bound states with chiral magnetic textures}

\author{Utkan G\"ung\"ord\"u}%
 \email{utkan@lps.umd.edu}
 \affiliation{Laboratory for Physical Sciences, College Park, Maryland 20740, USA}
 \affiliation{Department of Physics, University of Maryland, College Park, Maryland 20742, USA}
\author{Alexey A. Kovalev}
 \email{alexey.kovalev@unl.edu}
 \affiliation{Department of Physics and Astronomy and Nebraska Center for Materials and Nanoscience, University of Nebraska, Lincoln, Nebraska 68588, USA}
\date{\today}

\begin{abstract}
The aim of this Tutorial is to give a pedagogical introduction into realizations of Majorana fermions, usually termed as Majorana bound states (MBS), in condensed matter systems with magnetic textures. We begin by considering the Kitaev chain model of `spinless' fermions and show how two `half' fermions can appear at chain ends due to interactions. By considering this model and its two-dimensional generalization, we emphasize intricate relation between topological superconductivity and possible realizations of MBS. We further discuss how ‘spinless’ fermions can
be realized in more physical systems, e.g., by employing the
spin-momentum locking. Next, we demonstrate how magnetic textures can be used to induce synthetic or fictitious spin-orbit interactions, and, thus, stabilize MBS.
We describe a general approach that works for arbitrary textures and apply it to skyrmions. We show how MBS can be stabilized by elongated skyrmions, certain higher order skyrmions, and chains of skyrmions. We also discuss how braiding operations can be performed with MBS stabilized on magnetic skyrmions. This Tutorial is aimed at students at graduate level.
\end{abstract}

\maketitle

\section{Introduction}
The idea of Ettore Majorana~\cite{Majorana1937} that a fermion can be its own antiparticle found applications in high energy physics in proposals suggesting existence of Majorana fermions. Condensed matter systems comprised of large number of constituents with interactions can exhibit emergent excitations similar to Majorana fermions.~\cite{Kitaev2003,RevModPhys.80.1083,Alicea2012,Beenakker2013,RevModPhys.87.137} Defects in topological superconductors can host Majorana bound states (MBS).~\cite{RevModPhys.82.3045,Sato2017,Jiao2020} Topological superconductivity can be realized in heterostructures,~\cite{Aguado2020,Laubscher2021} and MBS have been predicted at edges or surfaces of topological insulators proximized with superconductors~\cite{PhysRevLett.100.096407,PhysRevLett.101.120403,PhysRevB.79.161408} as well as in semiconducting quantum wires and wells with similar proximity to superconductor.~\cite{PhysRevLett.105.077001,PhysRevLett.105.177002,PhysRevB.82.214509,PhysRevB.84.144522,PhysRevLett.104.040502,PhysRevB.81.125318,PhysRevB.82.134521} The presence of spin-orbit interaction and spin-momentum locking is crucial for these proposals. It is known that magnetic texture can lead to an effective spin-orbit interaction,~\cite{PhysRevB.82.045127,PhysRevX.3.011008} hence it is also possible to stabilize MBS by employing chains of magnetic adatoms deposited on a superconductor~\cite{PhysRevB.88.155420,PhysRevB.88.020407,PhysRevB.84.195442,PhysRevLett.111.186805,PhysRevLett.111.206802,PhysRevLett.111.147202} or by employing magnetic textures~\cite{PhysRevLett.109.236801,PhysRevB.93.224505,PhysRevB.94.214509,PhysRevB.97.115136,PhysRevB.100.064504,Garnier2019,PhysRevMaterials.4.081401,PhysRevB.104.214501,Desjardins2019,PhysRevApplied.12.034048,PhysRevLett.117.077002,MatosAbiague2017,PhysRevB.99.134505,PhysRevB.101.024514,Mohanta2021} in a proximity of superconductor.  

The above platforms often do not address the main challenge of finding unquestionable experimental proofs of MBS existence. The difficulty arises due to the fact that experimental signatures of MBS~\cite{Mourik2012,Das2012,Rokhinson2012,Deng2012,PhysRevLett.109.056803,PhysRevLett.109.186802,Deng2016} (e.g., measurements of zero-bias conductance peaks) often allow for an alternative explanation involving Andreev bound states or Yu-Shiba-Rusinov states.~\cite{PhysRevB.86.100503,PhysRevB.86.180503,PhysRevLett.109.186802,PhysRevB.96.075161,PhysRevB.96.195430,PhysRevB.98.235406,PhysRevB.97.165302,PhysRevB.98.245407,Prada2020,Kayyalha2020,PhysRevLett.120.156803,Valentini2021} As MBS are non-Abelian anyons appearing at zero energy separated from conventional excitations by an energy gap, they are suitable for manipulating and encoding quantum information.~\cite{RevModPhys.80.1083} It is clear that an experimental proof relying on non-Abelian statistics will provide the most definitive signature of MBS.~\cite{Brennen_2009,Alicea2011,Campbell2014,PhysRevB.90.115404,PhysRevB.94.235446,Plugge2017,PhysRevB.95.235305,Zhou2022} Experimental realizations will, however, require careful engineering that should also address the question of accurate MBS initialization.~\cite{PhysRevB.90.115404,Zhou2022} Other signatures that can potentially reveal the presence of MBS include the fractional Josephson effect.~\cite{Kitaev:2001,PhysRevLett.105.077001,PhysRevB.79.161408,PhysRevLett.107.236401,PhysRevLett.105.177002,PhysRevB.84.081304,PhysRevLett.108.257001,PhysRevLett.106.077003,PhysRevLett.107.177002,PhysRevB.86.140503,PhysRevB.86.140504} and current signatures in planar Josephson junctions.~\cite{PhysRevX.7.021032,PhysRevLett.126.036802,arxiv.2201.03453} In particular, the fractional Josephson effect reveals a current with periodicity $4\pi$ across a Josephson junction as opposed to the conventional $2\pi$ periodicity.

Stabilization of MBS with chiral magnetic textures such as skyrmions can be particularly beneficial due to robustness associated with topological protection of skyrmions.~\cite{PhysRevB.93.224505,PhysRevB.97.115136,PhysRevB.100.064504,Garnier2019,PhysRevMaterials.4.081401,PhysRevLett.126.117205,Mohanta2021,PhysRevB.104.214501} Skyrmions, initially observed in a form of skyrmion lattices in chiral magnets,~\cite{Rler2006,Muhlbauer2009,Yu2010} are under active investigation due to their useful properties for spintronics applications.~\cite{EverschorSitte2018}  Furthermore, skyrmions can be manipulated by  charge currents,~\cite{Sampaio2013,Iwasaki2013,PhysRevB.89.064425} surface acoustic
waves,~\cite{Nepal2018,Yokouchi2020} as wells as gradients of strain,~\cite{Yanes2019} magnetic field,~\cite{Zhang2018,PhysRevB.101.064408,Casiraghi2019} or temperature;~\cite{Kovalev2012,PhysRevB.89.241101,PhysRevLett.111.067203,Mochizuki2014,PhysRevLett.112.187203} thus, making it possible to probe non-Abelian statistics of MBS~\cite{PhysRevB.97.115136,PhysRevB.104.214501} through skyrmion manipulations.
\begin{figure}
\centering
\includegraphics[width=\linewidth]{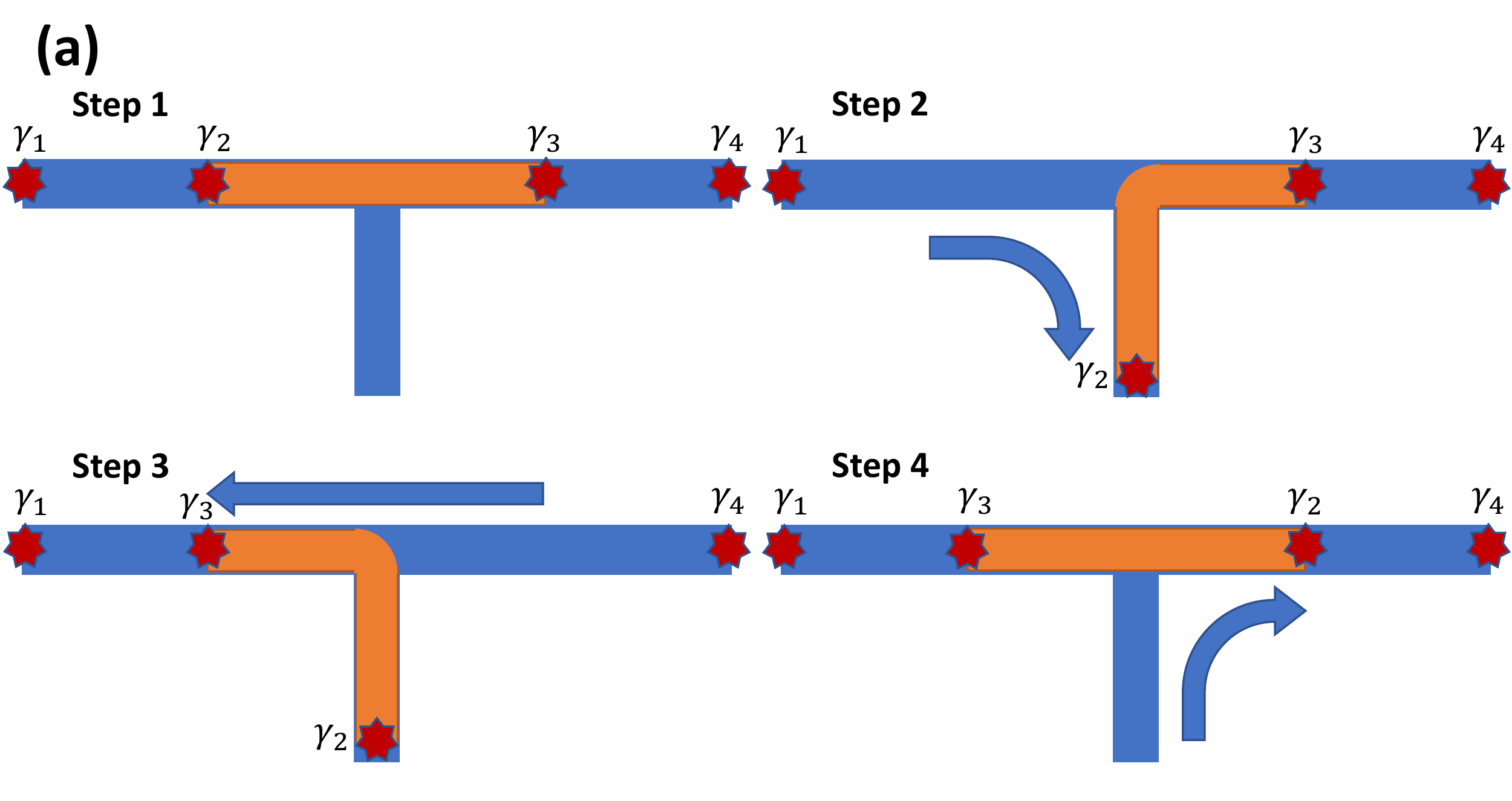}
\includegraphics[width=\linewidth]{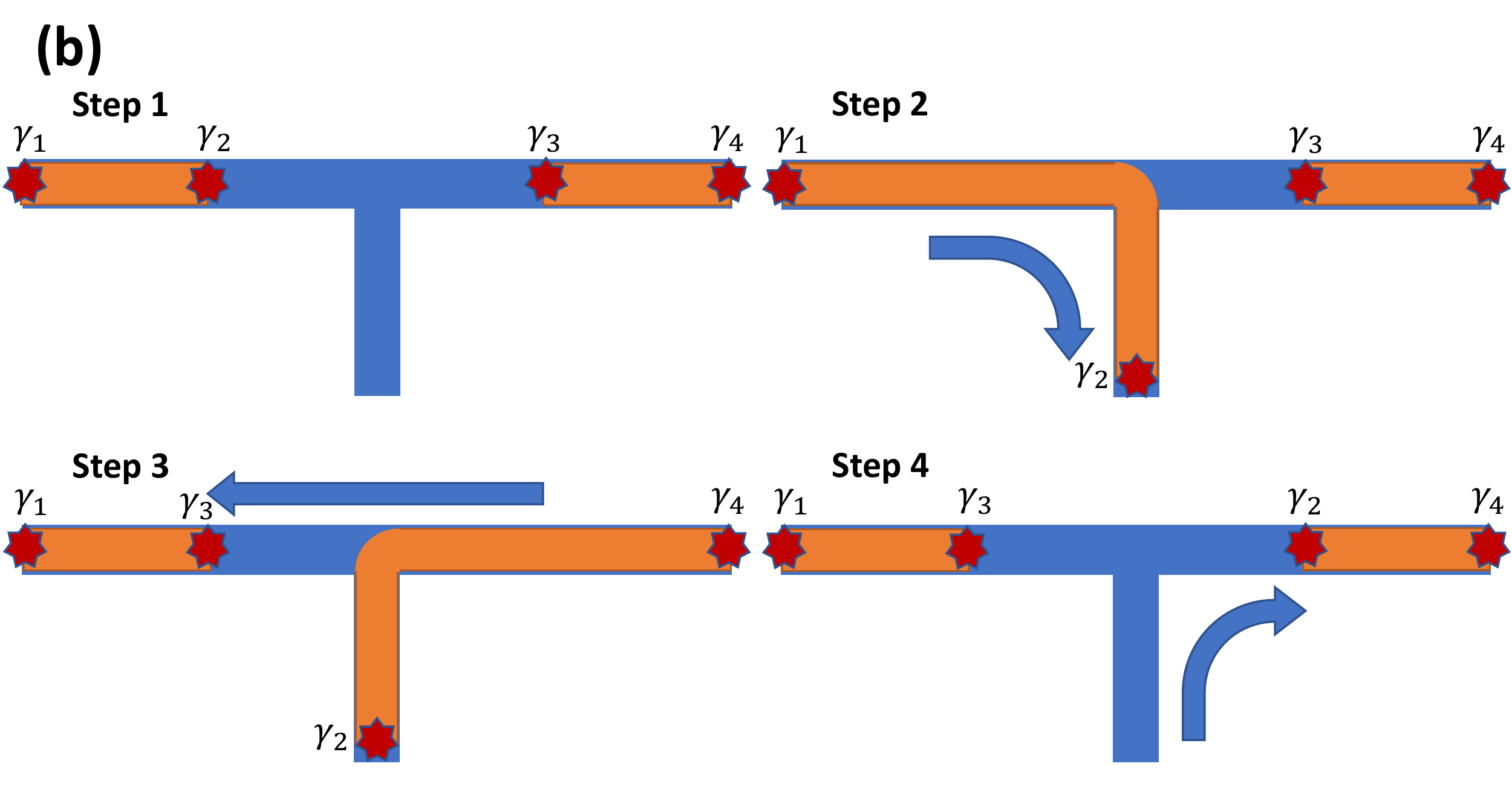}
\caption{An outline of steps realizing adiabatic exchange of MBS across topological (a) or trivial (b) regions in 1D or quasi 1D geometries. Arrows indicate directions of adiabatic motion of MBS bound to domain walls between topological and trivial regions.}
\label{fig:braiding}
\end{figure}

In many proposals, MBS are stabilized in 1D or quasi-1D geometries at domain walls between topological and trivial regions. Moving domain walls then leads to a natural way of braiding MBS; however, effective braiding can be also achieved without physically moving MBS via tunable couplings between MBS and measurements.~\cite{PhysRevB.95.235305} Such or similar experimental realizations of braiding can lead to direct confirmations of non-Abelian statistics. In Fig.~\ref{fig:braiding}, we sketch a possible proof of principle realization of braiding operation by  adiabatic exchange of MBS across topological or trivial regions.~\cite{Alicea2011}
It is important to note that in many MBS proposals, e.g., based on semiconductor nanowires, 1D or quasi-1D geometry does not prevent realization of non-Abelian statistics.~\cite{Alicea2011,PhysRevB.94.235446,Plugge2017,PhysRevB.95.235305} Furthermore, a practical realization may require a usage of Y- or X-shaped junctions to optimize the topological gap.~\cite{PhysRevB.89.245405,PhysRevB.90.115429,PhysRevLett.124.137001} 
In step 1, four MBS, $\gamma_{1}$, $\gamma_{2}$, $\gamma_{3}$, and $\gamma_{4}$, are stabilized at domain walls between topological and trivial regions. In step 2, by adiabatic motion of domain wall the mode $\gamma_2$ is shifted into auxiliary region. In step 3, the mode $\gamma_3$ is adiabatically placed in the initial position of mode $\gamma_2$. In step 4, the mode $\gamma_2$ is adiabatically placed in the initial position of mode $\gamma_3$. As the four MBS are fermion-like quasiparticles, they can be described by standard anticommutation relations, i.e., in the language of second quantized creation and annihilation operators, $\{{\gamma}_i,{\gamma}_j^\dagger \}=2\delta_{ij}$ where it is convenient to renormalize operators by introducing additional $2$ in the right hand side. Taking into account the fact that these quasiparticles are their own antiparticles, we also have a relation ${\gamma}_i={\gamma}_i^\dagger$ (as such, unlike fermionic operators, MBS operators obey $\gamma_i^2 = {\gamma_i^\dagger}^2 = 1$). With such notations, it is not hard to check that the braiding exchange operation between MBS $i$ and $j$ can be expressed by the unitary operator $\hat{U}_{ij}=(1+{\gamma}_i {\gamma}_j)/\sqrt{2}$.\cite{PhysRevLett.86.268} Alternatively, we can write the transformation of MBS operators under braiding, $\gamma_i\rightarrow \gamma_j$ and $\gamma_j\rightarrow -\gamma_i$. Following a different braiding order, we can adiabatically guide the degenerate ground state of our system into different many-body states with the same energy, as follows from the commutator $\left[\hat{U}_{ij},\hat{U}_{jk}\right]={\gamma}_i{\gamma}_k$. In a different language, braiding operations correspond to quantum gates performed on an array of MBS. For detailed discussions of quantum computations relying on braiding operations with MBS see Refs.~[\onlinecite{Alicea2012,Leijnse2012,Sarma2015}].

The Tutorial is organized as follows. In Sec.~II, we discuss a simple model, usually referred to as the Kitaev chain model of `spinless' fermions, realizing two MBS at its ends. By considering this model and its two-dimensional generalization, we emphasize the intricate relation between topological superconductivity and possible realizations of MBS in a pedagogical way. We show that in the topological superconducting state electrons effectively split into two `half' fermions, which can form MBS when spatially separated. We further discuss how `spinless' fermions can be realized in more physical systems, e.g., by employing the spin-momentum locking. We describe a well known realization based on a semiconductor nanowire with strong spin-orbit coupling proximized with an $s$-wave superconductor. By using a local unitary transformation, we show that spin-orbit interaction can be replaced by the synthetic or fictitious spin-orbit interaction induced by a texture of magnetic moments. This shows a potential utility of chiral magnetic textures, such as skyrmions, for realizations of MBS. In Sec.~III, we describe the model of a 2DEG that forms at the interface between an $s$-wave superconductor and a chiral ferromagnet hosting a magnetic texture, with interface induced Rashba spin-orbit coupling and subject to an external magnetic field. We discuss the conditions to obtain elongated skyrmions, as well as the advantages and limitations of using such skyrmions to stabilize MBS. In Sec.~IV, we numerically study the formation of MBS using three different magnetic textures: an elongated skyrmion, a chain of circular skyrmions, and a skyrmion with high ($>1$) azimuthal and radial magnetization flips. The elongated skyrmion and the skyrmion chain both effectively realize a quantum wire with localized MBS forming at their ends, whereas the skyrmion with high winding numbers leads to a single localized MBS at its core and a delocalized MBS outside. In Sec.~V, we outline methods to realize quantum operations through braiding of the MBS for elongated skyrmions and the skyrmion chain. Sec.~VI concludes the tutorial with a summary and an outlook of future directions.
\begin{figure}
\centering
\includegraphics[width=\linewidth]{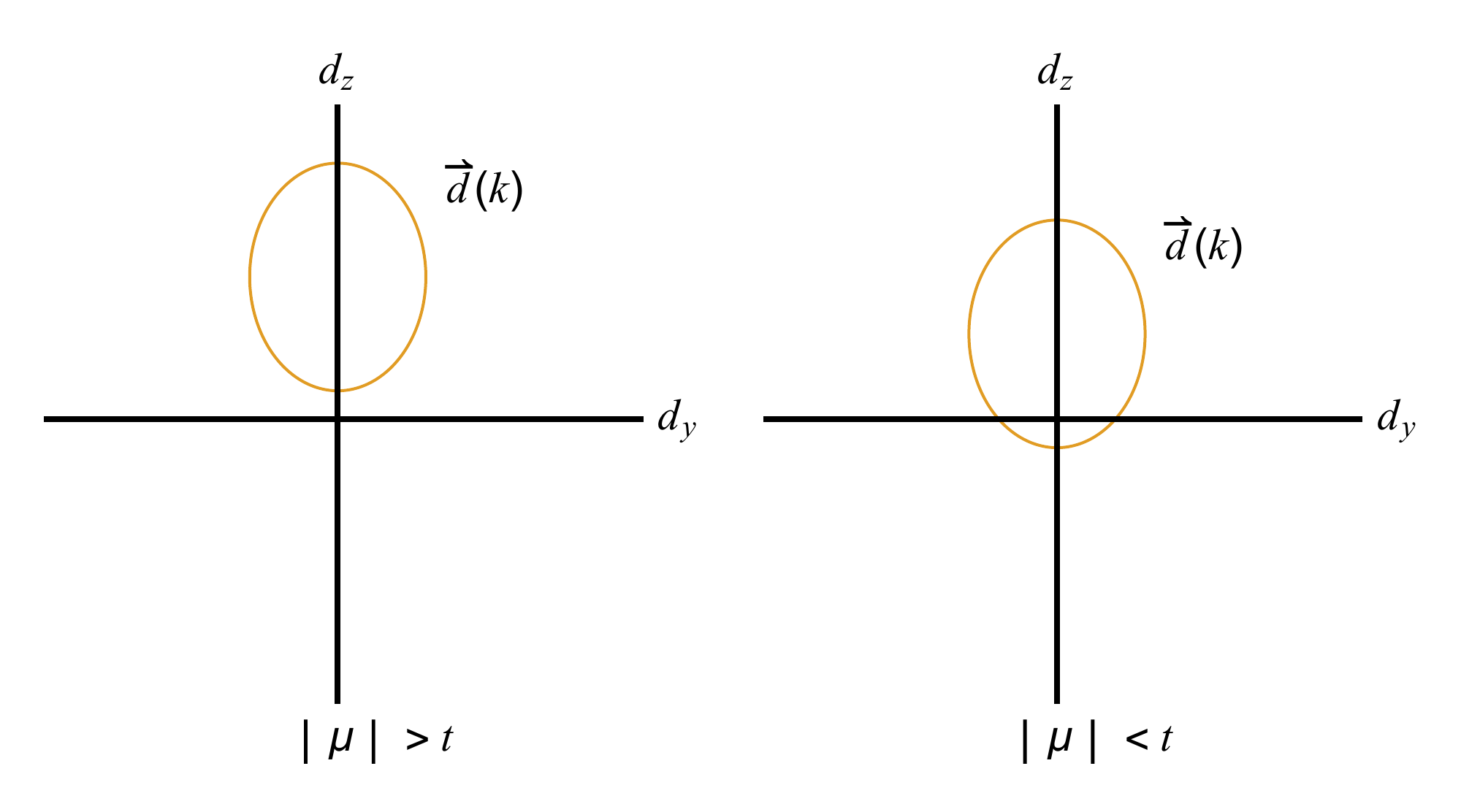}
\caption{The vector $\vec{d}(k)$ describes an ellipse centered at $d_z=-\mu$. The presence of the origin outside or inside of the loop distinguishes the trivial and topological phases. A strong pairing phase corresponding to trivial superconductor is realized when $|\mu|>t$. A weak pairing phase corresponding to topological superconductor is realized when $|\mu|<t$.}
\label{fig:top}
\end{figure}
\section{A primer on Majorana bound states}
In this section, we review basic ideas that led to proposals of MBS in condensed matter systems. Superconductors described by Bogoliubov de Gennes (BdG)
Hamiltonians appear to be useful for realizing MBS described by operators with the property ${\gamma}_i={\gamma}_i^\dagger$ due to the particle-hole symmetry. However, a usual $s$-wave superconductor cannot be used for realizing MBS due to the presence of spin index in BdG transformation, as in this case the particle-hole symmetry does not guarantee the presence of MBS at zero energy. A spinless $p$-wave superconductor circumvents this problem. A model of spinless $p$-wave superconductor can be realized in a chain of $N$ spinless fermions described by the Hamiltonian
\begin{equation}\label{eq:chain}
H=\sum\limits_{j=1}^{N-1} \left(-\frac{t}{2} c_j^\dagger c_{j+1}+\Delta c_j c_{j+1}+\text{H.c.}\right)-\sum\limits_{j=1}^{N}\mu c_{j}^\dagger c_j,
\end{equation}
where $t$ is the parameter describing hopping, $\Delta$ describes the superconducting pairing, and $\mu$ is the chemical potential. After applying the Fourier transform to Eq.~\eqref{eq:chain}, we obtain a bulk Hamiltonian
\begin{equation}
    H=\frac{1}{2}\sum\limits_{k}\begin{pmatrix}c_k^\dagger,  \,c_{-k}\end{pmatrix}H_\text{BdG}(k)\begin{pmatrix}c_k\\c_{-k}^\dagger\end{pmatrix},
\end{equation}
where $H_\text{BdG}(k)=\vec{d}(k)\cdot \vec{\tau}$, $\vec{\tau}$ is a vector of Pauli matrices describing the particle-hole sector, and  $\vec{d}(k)=(0,\Delta \sin k,-t \cos k-\mu)$. By plotting the vector $\vec{d}(k)$ in the $y-z$ plane as $k$ sweeps over the entire Brillouin zone, we can distinguish a trivial and topological supeconductor phases. As shown in Fig.~\ref{fig:top}, the vector $\vec{d}(k)$ describes an ellipse centered at $d_z=-\mu$. The presence of the origin outside or inside of the ellipse distinguishes the trivial and topological phases, respectively.~\cite{PhysRevLett.109.150408} Note that the $\mathbb{Z}_2$ Pfaffian invariant of a 1D wire corresponds to the parity of the winding number associated with $\vec{d}(k)$. In particular, the value $-1$ of the $\mathbb{Z}_2$ invariant corresponds to the topological phase in Fig.~\ref{fig:top} and the value $1$ of the $\mathbb{Z}_2$ invariant corresponds to the trivial phase in Fig.~\ref{fig:top}.
\begin{figure}
\centering
\includegraphics[width=\linewidth]{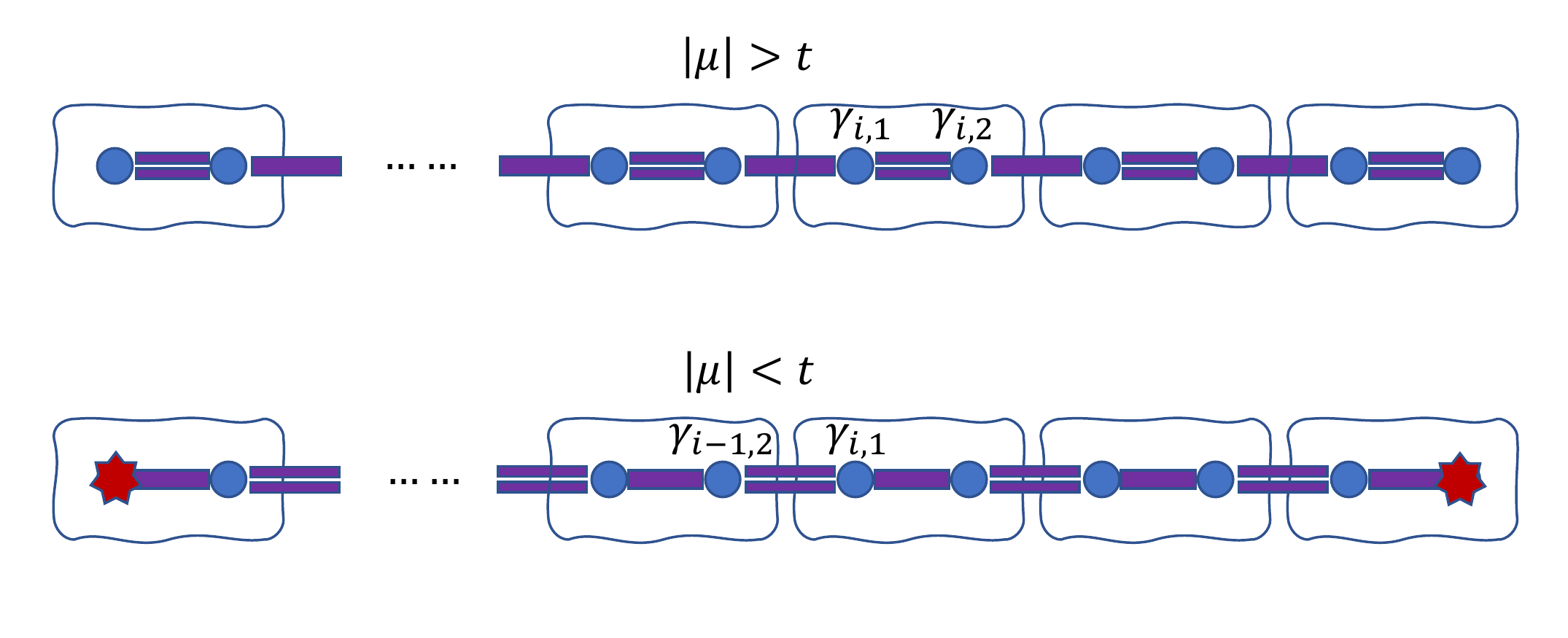}
\caption{The Majorana operator representation of the fermion chain. For $|\mu|>t$, the pairing of odd bonds  dominates leading to a trivial superconductor state without MBS. For $|\mu|<t$, the pairing of even bonds dominates leading to a topological superconductor state with MBS at the ends of the chain.}
\label{fig:chain}
\end{figure}

We can further unravel the significance of topological superconductivity by switching to Majorana operators in the Hamiltonian in Eq.~\eqref{eq:chain}. We introduce operators $\gamma_{i,1}$ and $\gamma_{i,2}$ according to relations
\begin{equation}
    \gamma_{i,1}=c_i^\dagger+c_i,\,\, \gamma_{i,2}=i(c_i^\dagger-c_i),
\end{equation}
which can be interpreted as the real and imaginary parts of the electron annihilation operator. The new operators satisfy the anticommutation relations of Majorana operators,
\begin{equation}
    \{\gamma_{i,\alpha},\gamma_{i,\beta}\}=2\delta_{ij}\delta_{\alpha\beta},\,\,\gamma_{i,\alpha}=\gamma_{i,\alpha}^\dagger.
\end{equation}
In terms of Majorana operators, the Hamiltonian in Eq.~\eqref{eq:chain} takes the form
\begin{align}\nonumber
    H=&\frac{i}{2}\sum\limits_{j=1}^{N-1} \left[(\Delta -t/2) \gamma_{j,1} \gamma_{j+1,2}+(\Delta +t/2)\gamma_{j,2} \gamma_{j+1,1}\right]\\
    &-\frac{\mu}{2}\sum\limits_{j=1}^{N}(i\gamma_{j,1} \gamma_{j,2}+1).\label{eq:mop}
\end{align}
An easy way to see the appearance of MBS is by limiting the consideration to the case $\Delta=t/2$. Up to an unimportant constant, we obtain the Hamiltonian with two types of pairing terms,
\begin{equation}\label{eq:Mchain}
    H=\frac{t}{2}\sum\limits_{j=1}^{N-1} i \gamma_{j,2} \gamma_{j+1,1}
    -\frac{\mu}{2}\sum\limits_{j=1}^{N}i\gamma_{j,1} \gamma_{j,2},
\end{equation}
by using the Majorana operator representation of the fermion chain. We observe that for $|\mu|>t$ the pairing of odd bonds in Eq.~\eqref{eq:Mchain} dominates leading to a trivial superconductor state without MBS, and for $|\mu|<t$ the pairing of even bonds in Eq.~\eqref{eq:Mchain} dominates leading to a topological superconductor state with MBS at the ends of the chain, see Fig.~\ref{fig:chain}. 
\begin{figure}
\centering
\includegraphics[width=\linewidth]{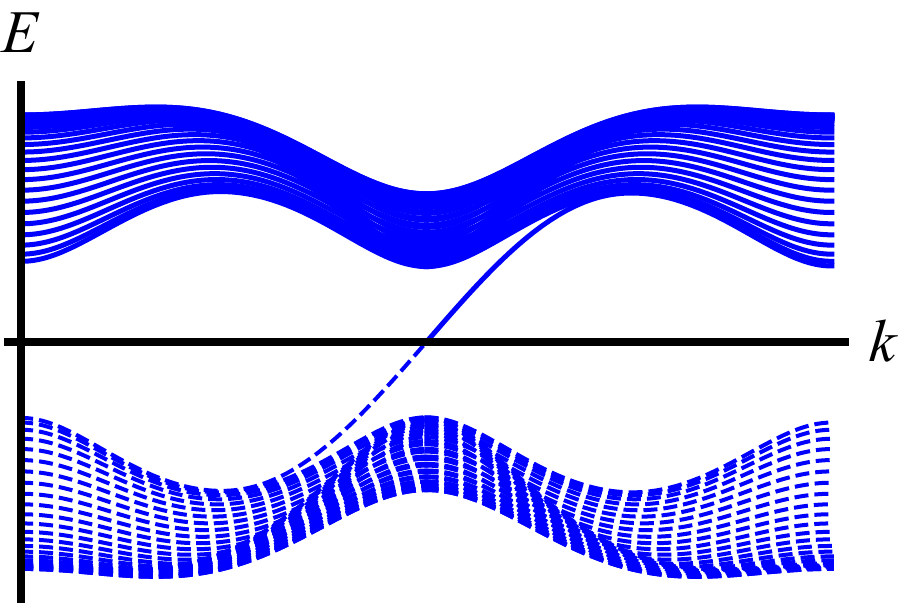}
\caption{The energy spectrum as a function of angular momentum along the strip boundary for a topological superconductor in Fig.~\ref{fig:vortex}(a) described by the Hamiltonian in Eq.~\eqref{eq:chain2D}. The energy spectrum reveals a chiral Majorana edge mode in the superconducting gap.}
\label{fig:strip}
\end{figure}

\begin{figure}
\centering
\includegraphics[width=0.8\linewidth]{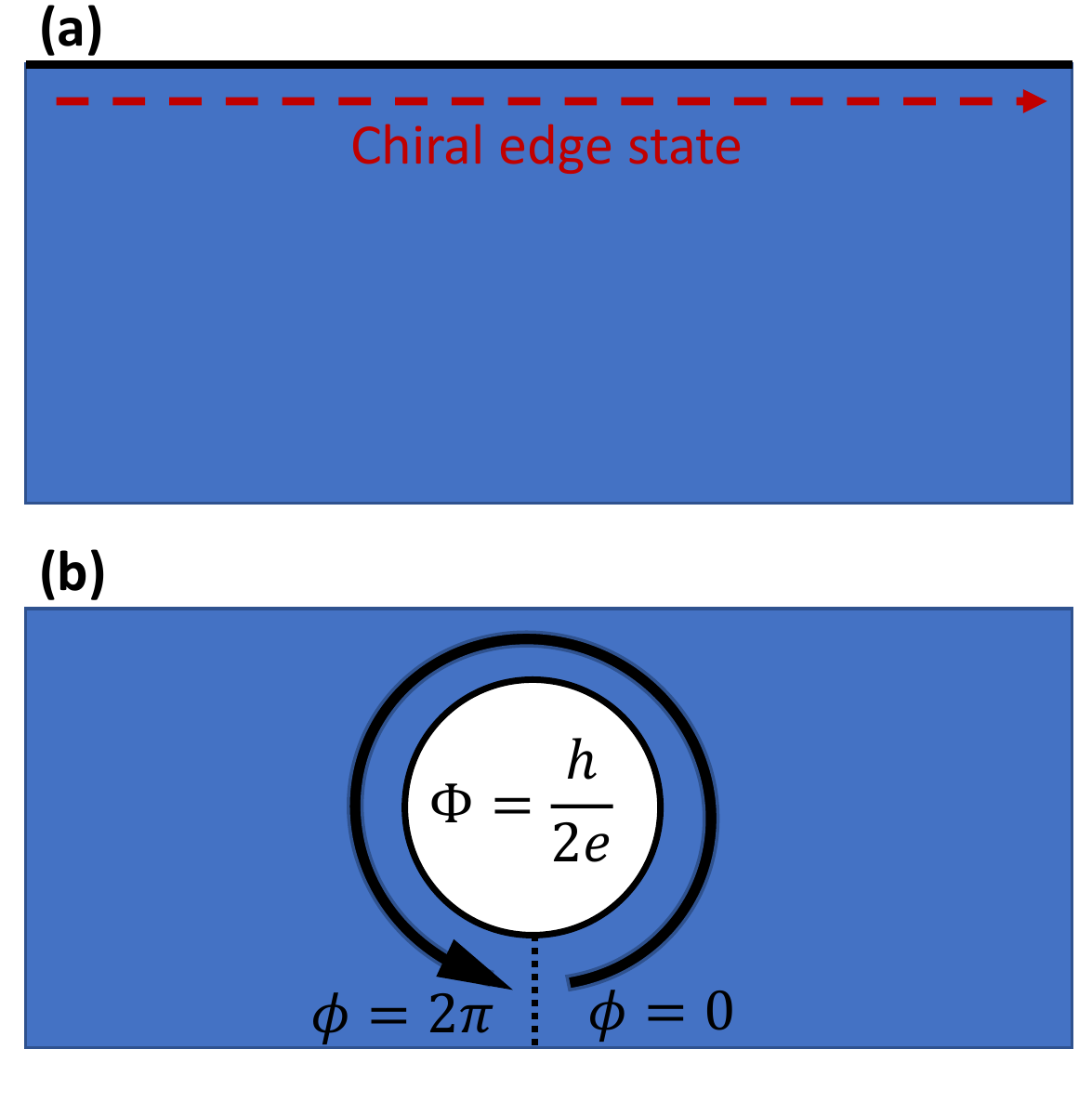}
\caption{ (a) A boundary between topological and trivial regions contains a chiral Majorana edge mode. (b) Majorana edge states localized on a circular boundary between topological and trivial regions will have half integer angular momentum and nonzero energy due to anti-periodic boundary conditions. A flux $\Phi=\frac{h}{2e}$ introduces a branch cut and changes boundary conditions to periodic, which leads to a localized Majorana state at zero energy.}
\label{fig:vortex}
\end{figure}

The chain model in Eq.~\eqref{eq:chain} can be generalized to two-dimensional model of topological superconductor on a square lattice with the Hamiltonian~\cite{PhysRevLett.105.227003}
\begin{align}\nonumber
H=&\sum\limits_{\mathbf{r}} (-\frac{t}{2} c_{\mathbf{r}}^\dagger c_{\mathbf{r}+\boldsymbol\delta_x}-\frac{t}{2}c_{\mathbf{r}}^\dagger c_{\mathbf{r}+\boldsymbol\delta_y}+i\Delta c_{\mathbf{r}} c_{\mathbf{r}+\boldsymbol\delta_x}+\Delta c_{\mathbf{r}} c_{\mathbf{r}+\boldsymbol\delta_y}\\&+\text{H.c.})-\sum\limits_{\mathbf{r}}\mu c_\mathbf{r}^\dagger c_\mathbf{r}\,,\label{eq:chain2D}
\end{align}
where $\boldsymbol\delta_x$ and $\boldsymbol\delta_y$ describe bonds along the $x$-axis and $y$-axis, respectively. After applying the Fourier transform to Eq.~\eqref{eq:chain2D}, we obtain the bulk Bogoliubov–de Gennes Hamiltonian
\begin{equation}
    H=\frac{1}{2}\sum\limits_{\mathbf{k}}\begin{pmatrix}c_\mathbf{k}^\dagger  ,\,c_{-\mathbf{k}}\end{pmatrix}H_\text{BdG}(\mathbf{k})\begin{pmatrix}c_\mathbf{k}\\c_{-\mathbf{k}}^\dagger\end{pmatrix},
\end{equation}
where $H_\text{BdG}(\mathbf{k})=\vec{d}(\mathbf{k})\cdot \vec{\tau}$, $\vec{\tau}$ is a vector of Pauli matrices describing the particle-hole sector, and  $\vec{d}(\mathbf{k})=(\Delta \sin k_x,\Delta \sin k_y,-t (\cos k_x+ \cos k_y)-\mu)$. The vector $\vec{d}(\mathbf{k})/|\vec{d}(\mathbf{k})|$ maps the Brillouin zone of the two-dimensional lattice to the 2-sphere of the three-component unit vector, which can be described by the homotopy group $\pi_2(S^2)$ and topological invariant $\mathbb{Z}$.~\cite{PhysRevB.61.10267} A Chern number determines the number of Majorana chiral edge modes at the boundary with a trivial state. By calculating the topological invariant $\mathbb{Z}$ for the model in Eq.~\eqref{eq:chain2D}, one can identify that the model with $|\mu|>2t$ realizes the strong pairing phase of a trivial superconductor, and the model with $|\mu|<2t$ realizes the weak pairing phase of a topological superconductor. At a boundary between the trivial and topological phases there exists a chiral Majorana edge mode, as shown in Figs.~\ref{fig:strip} and \ref{fig:vortex}. Such chiral Majorana edge modes can be used for realizations of localized MBS at zero energy in the presence of a magnetic flux $\Phi=\frac{h}{2e}$ and a vortex;\cite{Volovik1999} see Fig.~\ref{fig:vortex}(b), where the vortex core corresponds to the trivial region.~\cite{Alicea2012} Without a magnetic flux in Fig.~\ref{fig:vortex}(b), Majorana edge states localized on a boundary between topological and trivial regions will have half integer angular momentum and nonzero energy due to anti-periodic boundary conditions. A flux $\Phi=\frac{h}{2e}$ introduces a branch cut and changes boundary conditions to periodic, which leads to a localized MBS at zero energy.
Note that the particle-hole symmetry requires MBS to form in pairs, and as a result the second MBS will form on another vortex or the system boundary in Fig.~\ref{fig:vortex}(b) (the second MBS is not shown in Fig.~\ref{fig:vortex}).

To realize MBS, we have assumed the presence of spinless fermions and superconducting pairing. As $p$-wave superconductors are rare, one may need to resort to heterostructures for practical realizations of MBS. To achieve superconducting pairing, one can use an ordinary $s$-wave superconductor proximized with non-superconducting system.~\cite{Zutic2019} Furthermore, the spin degree of freedom has to be frozen in non-superconducting system, which can be achieved in the presence of strong spin-orbit interaction and spin-momentum locking. To this end, MBS have been predicted at edges or surfaces of topological insulators proximized with superconductors~\cite{PhysRevLett.100.096407,PhysRevLett.101.120403,PhysRevB.79.161408} as well as in semiconducting quantum wires and wells with similar proximity to superconductor.~\cite{PhysRevLett.105.077001,PhysRevLett.105.177002,PhysRevB.82.214509,PhysRevB.84.144522,PhysRevLett.104.040502,PhysRevB.81.125318,PhysRevB.82.134521} 
\begin{figure}
\centering
\includegraphics[width=0.8\linewidth]{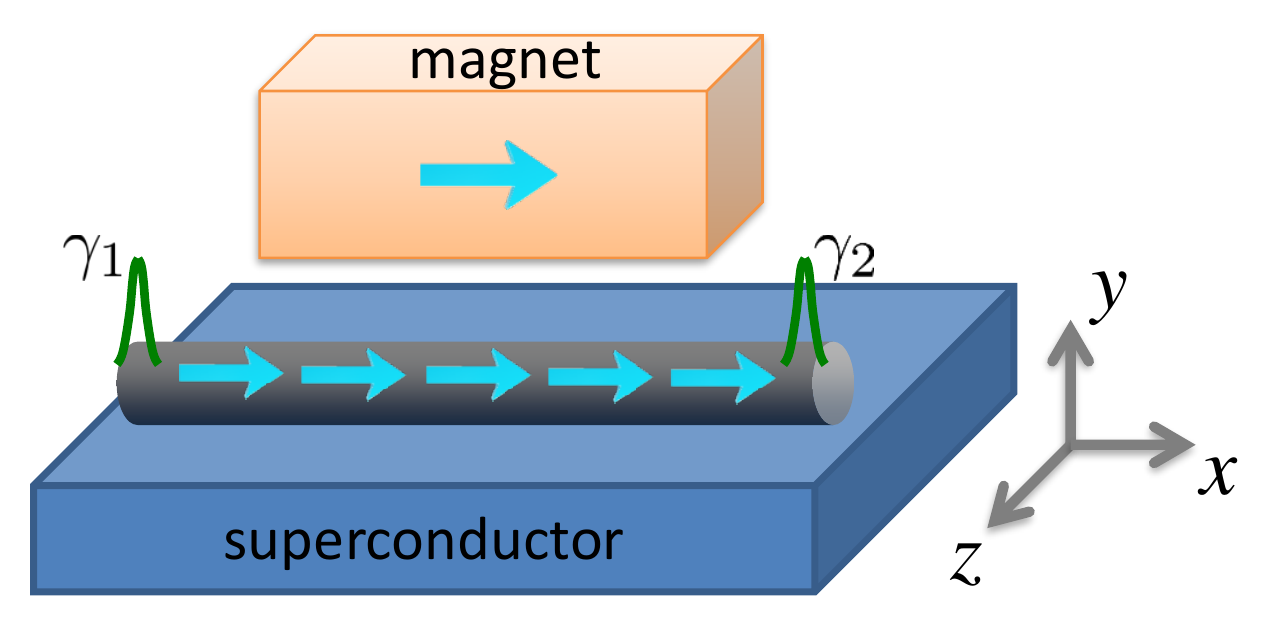}
\caption{A 1D semiconductor nanowire with strong
spin-orbit coupling is placed on top of an $s$-wave superconductor. In the presence of external magnetic field, Majorana bound states can be formed by sharp profile of stray magnetic fields of a nanomagnet.~\cite{PhysRevLett.112.106402,PhysRevLett.117.077002} }
\label{fig:wire}
\end{figure}

\begin{figure}
\centering
\includegraphics[width=0.8\linewidth]{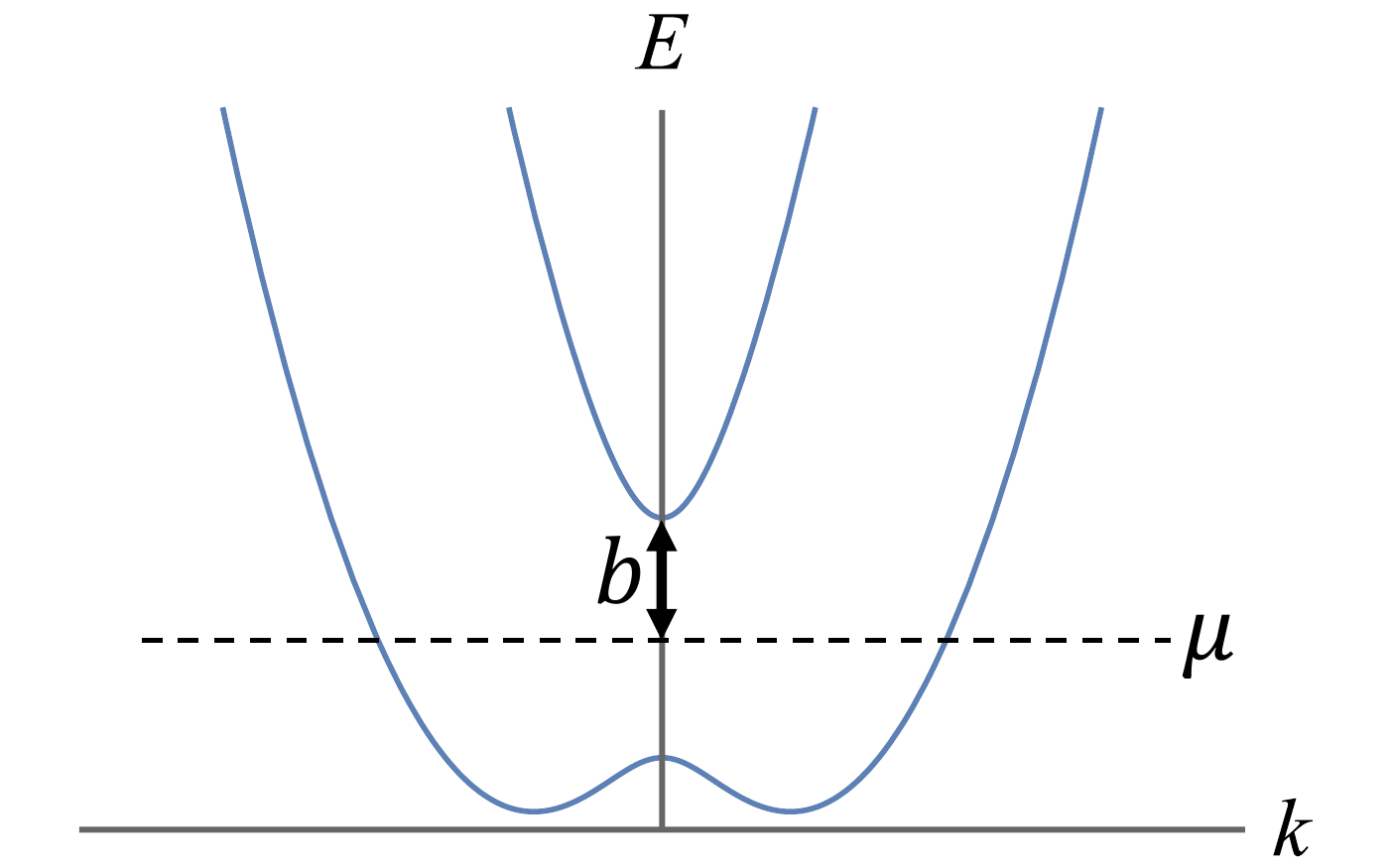}
\caption{Spectrum of a 1D semiconductor nanowire with strong
spin-orbit coupling. The spectrum is split horizontally by spin-orbit interaction, and a gap opens at zero momentum due to the presence of magnetic field. }
\label{fig:spec}
\end{figure}

A possible realization based on a 1D semiconductor nanowire proximized with an $s$-wave superconductor is shown in Fig.~\ref{fig:wire}. In the presence of spin-orbit interaction and uniform magnetic field, the wire can be described by the Hamiltonian
\begin{equation}\label{eq:socwire}
    H_\text{wire}(k)=\frac{\hbar^2 k^2}{2 m^*}+\alpha k \sigma_z +\mathbf{b} \cdot \boldsymbol\sigma-\mu,
\end{equation}
where $m^*$ is the effective mass, $\alpha$ is the strength of the spin-orbit interaction, $\sigma_i$ are Pauli matrices describing the spin, and $\mathbf{b}$ describes an applied magnetic field along the $x$-axis, see Fig.~\ref{fig:wire}. The presence of magnetic field leads to opening of the gap in the spectrum of the Hamiltonian in Eq.~\eqref{eq:socwire} at zero momentum, which leaves only two modes within the gap, see Fig.~\ref{fig:spec}. This effectively realizes `spinless' fermions given that the chemical potential is properly tuned. The $s$-wave superconductor introduces pairing terms in the Hamiltonian in Eq.~\eqref{eq:socwire}, resulting in the Bogoliubov–de Gennes Hamiltonian
\begin{equation}
    H=\frac{1}{2}\sum\limits_{k}\begin{pmatrix}c_{k\uparrow}^\dagger , c_{k\downarrow}^\dagger,c_{-k\downarrow},-c_{-k\uparrow}\end{pmatrix}H_\text{BdG}(k)\begin{pmatrix}c_{k\uparrow}\\c_{k\downarrow}\\c_{-k\downarrow}^\dagger\\-c_{-k\uparrow}^\dagger\end{pmatrix},
\end{equation}
with 
\begin{equation}\label{eq:bdgwire}
    H_\text{BdG}(k)=\frac{\hbar^2 k^2}{2 m^*}\tau_z+\alpha k \tau_z \sigma_z +\mathbf{b} \cdot \boldsymbol\sigma+\Delta \tau_x-\mu \tau_z,
\end{equation}
where $\Delta$ describes the superconducting pairing. The Hamiltonian in Eq.~\eqref{eq:bdgwire} realizes a topological phase for $b^2>\Delta^2+\mu^2$ and a trivial phase for $b^2<\Delta^2+\mu^2$. For $b^2=\Delta^2+\mu^2$ gap closes and the system undergoes a quantum phase transition. An interesting situation arises for nonuniform magnetic fields, where one can tune the system in and out of the topological phase by varying magnetic fields. In this case, the lattice version of Eq.~\eqref{eq:bdgwire} has to be considered to capture the effect of nonuniform magnetic fields.~\cite{PhysRevLett.112.106402,PhysRevLett.117.077002}

\begin{figure}
\centering
\includegraphics[width=\linewidth]{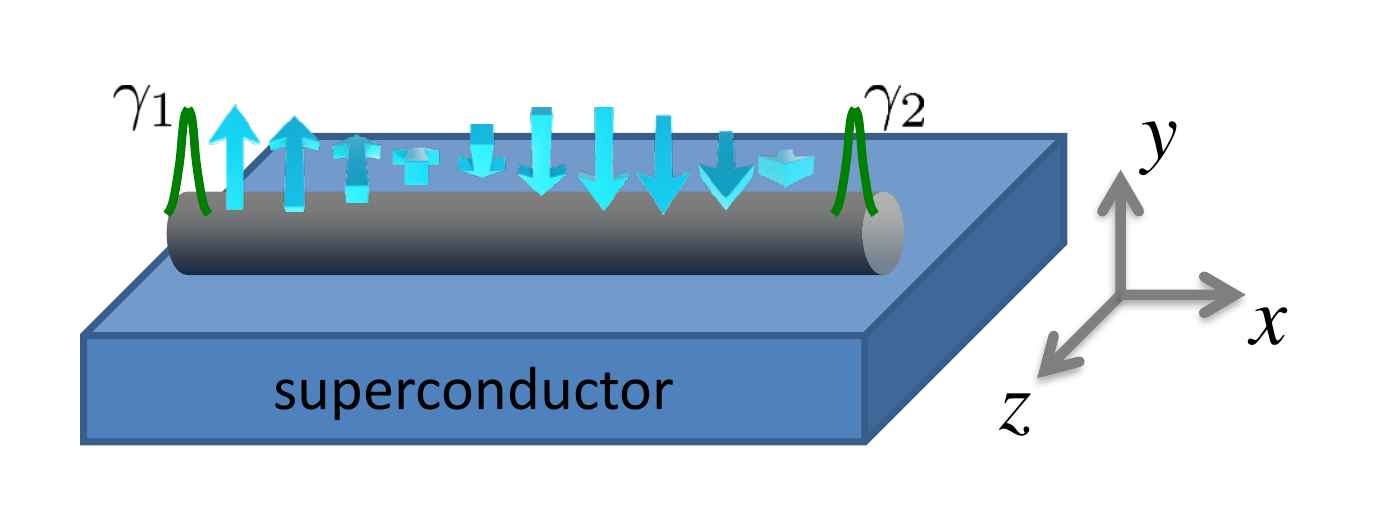}
\caption{A 1D semiconductor nanowire with strong
spin-orbit coupling is placed on top of an $s$-wave superconductor. A fictitious spin-orbit interaction can be induced through coupling to localized magnetic moments forming a magnetic helix. }
\label{fig:helix}
\end{figure}

It is possible to induce an effective spin-orbit interaction in 1D nanowire with a magnetic texture,~\cite{PhysRevB.82.045127,PhysRevX.3.011008,PhysRevLett.109.236801,PhysRevLett.111.186805,PhysRevLett.111.206802,PhysRevLett.111.147202} e.g., originating from a magnetic helix, see Fig.~\ref{fig:helix}. In Eq.~\eqref{eq:bdgwire}, we consider a static magnetic texture, $\mathbf{b}(x)=b[0,\cos(k_s x),\sin(k_s x)]^T$, in proximity to a 1D nanowire without spin-orbit interaction, i.e., $\alpha=0$, where $b$ and $k_s$ are the amplitude and wave vector of the helix. We assume that the magnetic texture helix is
coupled to the conductor via exchange interaction or stray field, and that the itinerant electrons in 1D nanowire do not affect the magnetic texture.
We can effectively rotate this helix back to collinear configuration by applying a local unitary transformation $U=e^{i\sigma_x k_s x/2}$ to the Hamiltonian, i.e., $H \rightarrow U H U^\dagger$, which introduces a vector potential in the transformation of momentum, $k\rightarrow k-i U \partial_x U^\dagger$. The transformed Hamiltonian can be written as    
\begin{equation}\label{eq:bdggauge}
    H_\text{BdG}(k)=\frac{\hbar^2 }{2 m^*}\tau_z\left(k-\frac{k_s}{2} \sigma_x\right)^2 +b \sigma_y+\Delta \tau_x-\mu \tau_z.
\end{equation}
The Hamiltonian in Eq.~\eqref{eq:bdggauge} is equivalent to Eq.~\eqref{eq:bdgwire} up to a rotation in spin space, where $\frac{\hbar^2 k_s}{2 m^*}$ plays a role of $\alpha$. The Hamiltonian in Eq.~\eqref{eq:bdggauge} realizes a topological phase for $b^2>\Delta^2+\left(\mu-\frac{\hbar^2 k_s^2}{8 m^*}\right)^2$ and a trivial phase for $b^2<\Delta^2+\left(\mu-\frac{\hbar^2 k_s^2}{8 m^*}\right)^2$. 

We note that many proposals of topological superconductors in heterostructures rely on the superconducting proximity.~\cite{Tewari2011,PhysRevLett.119.176805,Kotetes2013,Hendrickx2018,Ridderbos2019,PhysRevResearch.3.L022005,Zutic2019} Thus, careful engineering of heterostructures, e.g., by employing an insulating layer between semiconductor and superconductor, is important. In particular, disorder effects can lead to weak proximity and "soft" gap,~\cite{PhysRevLett.110.186803,Zhang2019} while strong hybridization can prevent realization of topological superconductivity.~\cite{PhysRevB.97.165425,PhysRevB.100.035426} 
For more comprehensive discussions of topological superconductivity and its realizations, we refer the reader to reviews.~\cite{RevModPhys.87.137,Alicea2012,Stanescu2013,Beenakker2013,Sato2016,Sato2017,Cimento2017,Lutchyn2018}
MBS are also susceptible to `quasiparticle poisoning' due to processes that do not conserve the fermion parity.~\cite{PhysRevB.85.174533} A number of solutions, e.g., relying on large charging energy, have been proposed.~\cite{PhysRevLett.116.050501,PhysRevB.94.174514,PhysRevB.94.235446,Plugge2017,PhysRevB.95.235305}

We mention in passing that the above ideas of using topological superconductivity for realizations of MBS can be further generalized by employing higher order topological insulators and superconductors. Such an $n$th-order topological insulator or superconductor exhibits $(d-n)$-dimensional boundary in $d$-dimensions. A two dimensional higher order topological superconductor can be used to realize MBS for $n=2$.~\cite{PhysRevB.96.060505,PhysRevB.97.205136,PhysRevB.98.165144,PhysRevB.98.245413,PhysRevB.97.205134,PhysRevB.100.075415,PhysRevResearch.2.043155,PhysRevB.105.195149,PhysRevLett.123.167001,PhysRevLett.122.126402,PhysRevLett.122.236401} 

\section{Majorana bound states and magnetic textures}
In this section, we describe how a general magnetic texture can affect two-dimensional electron gas (2DEG) in proximity to $s$-wave superconductor. We show how such a system can form regions with topological superconductivity.
\begin{figure}
\centering
\includegraphics[width=0.7\linewidth]{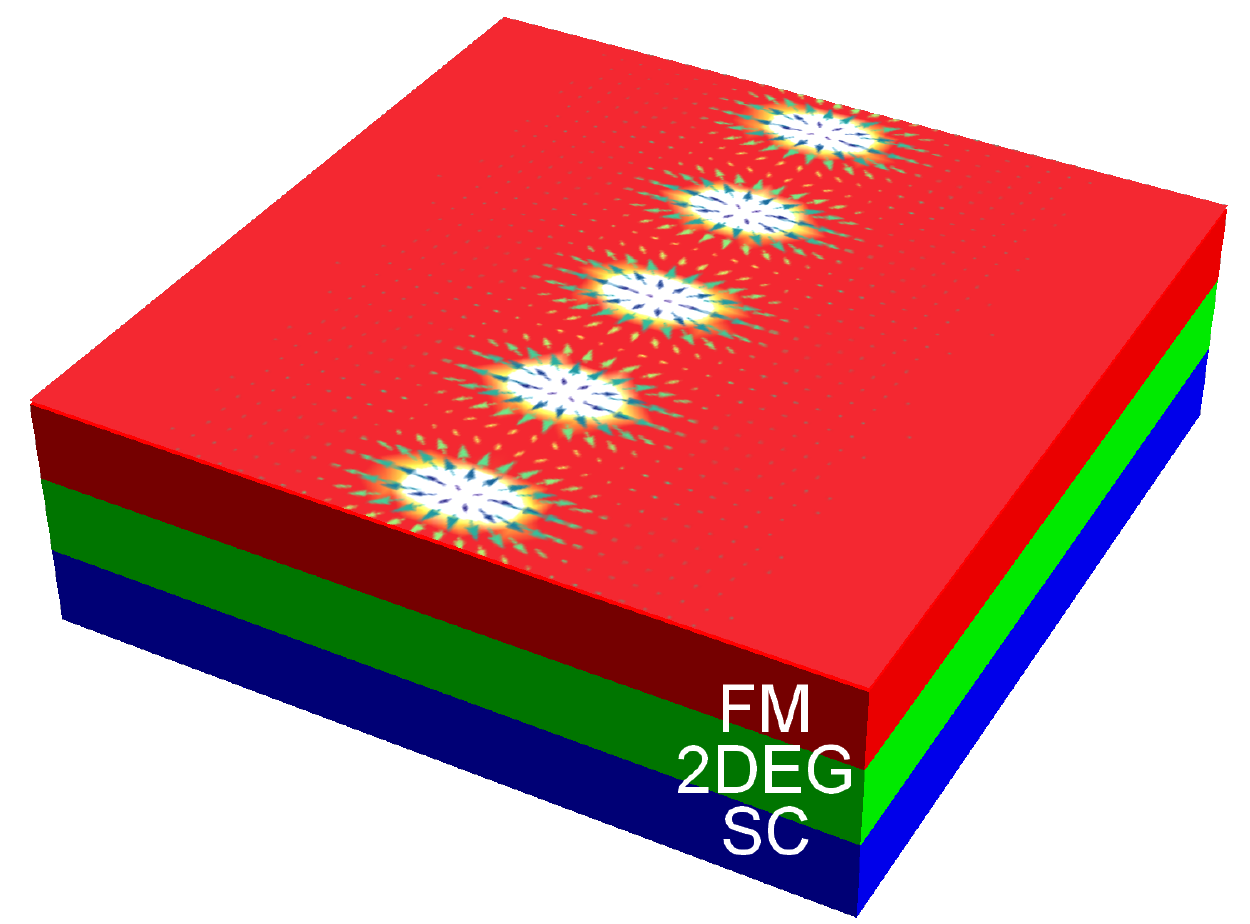}
\caption{A sketch of a possible device geometry. An $s$-wave superconductor (blue) is placed in proximity to 2DEG (green). A ferromagnet, antiferromagnet, or ferrimagnet (red) with magnetic texture is placed on top of 2DEG.}
\label{fig:2d-setup}
\end{figure}
We consider a 2DEG that is formed between an $s$-wave superconductor and a ferromagnet (in principle, it can also be a ferrimagnet or an antiferromagnet) hosting a non-uniform magnetic texture in the presence of an applied uniform magnetic field $B$ perpendicular to its plane. A sketch of possible device geometry is shown in Fig.~\ref{fig:2d-setup}. We model this system starting from the BdG Hamiltonian $H = \int d^2 r \Psi^\dagger H_\text{BdG} \Psi$ with
\begin{align}
H_\text{BdG} = \begin{pmatrix}
H_0 & \Delta e^{i\varphi} \\
\Delta e^{-i\varphi} & -\sigma_y H_0^* \sigma_y
\end{pmatrix}
\end{align}
with the single-particle Hamiltonian $H_0$ of the 2DEG given by
\begin{align}
H_0 = \frac{p^2}{2 m^*} - \mu - \frac{\alpha_R}{\hbar} (\boldsymbol e_z \times \boldsymbol p) \cdot \boldsymbol \sigma + \frac{1}{2}g \mu_B B \sigma_z - J \boldsymbol n \cdot \boldsymbol \sigma,
\end{align}
resulting in
\begin{align}
H_\text{BdG} = & \left[ \frac{p^2}{2 m^*} - \mu - \frac{\alpha_R}{\hbar} (\boldsymbol e_z \times \boldsymbol p) \cdot \boldsymbol \sigma \right] \tau_z + \nonumber\\
& \Delta [\cos\varphi \tau_x - \sin\varphi \tau_y] + \frac{1}{2}g^* \mu_B B \sigma_z - J \boldsymbol n \cdot \boldsymbol \sigma
\label{eq:H}
\end{align}
acting on the Nambu spinor $\Psi = (\psi_\uparrow^\dagger, \psi_\downarrow^\dagger, \psi_\downarrow, -\psi_\uparrow)$. Above, $m^*$ and $g^*$ are the effective mass and $g$-factors, $\boldsymbol p = -i \hbar \boldsymbol \nabla$, $\alpha_R$ is the strength of the Rashba spin-orbit coupling (SOC) caused by the broken inversion symmetry, $\mu$ is the chemical potential and $\mu_B$ is the Bohr magneton. $\Delta e^{i \varphi}$ is the mean-field superconducting pairing potential $\langle \psi^\dagger_\uparrow \psi^\dagger_\downarrow \rangle$ induced in the 2DEG by the superconductor, and as such, is weaker compared to its value within the bulk superconductor. At any given point in the plane, the orientation of the magnetic texture is given by the unit vector $\boldsymbol n = \boldsymbol n(x,y)$, and $J$ denotes the strength of the exchange field induced by the proximity of the ferromagnet which favors the alignment of the spins in the 2DEG with the magnetic texture.  Finally, $\tau_i$ and $\sigma_i$ are Pauli matrices acting on the particle-hole and spin spaces respectively, and $\boldsymbol e_z$ is the unit vector along the $z$-axis. The same model also applies to a proximitized semiconductor wire covered by the ferromagnet.~\cite{Kim2015b} We remark that we do not take into account dipolar fields produced by magnetic textures assuming that exchange interaction is dominant. This should work well for sharp textures producing weak dipolar fields and systems with small saturation magnetization, such as ferrimagnets. We note that dipolar fields produced by skyrmions can lead to formation of vortices in superconductor (or lead to skyrmion-vortex interaction).~\cite{Hals2016,PhysRevLett.122.097001,PhysRevB.103.174519,PhysRevLett.126.117205}

A nonuniform magnetic texture itself provides an effective spin-orbit coupling, which can be seen as follows. Let us define $\boldsymbol M \equiv -J \boldsymbol n + g^* \mu_B B \boldsymbol e_z/2$ as the overall ``magnetic texture", and parameterize its orientation using spherical coordinates $\boldsymbol M = M(\sin M_\theta \cos M_\phi, \sin M_\theta \sin M_\phi, \cos M_\theta) =  R_z(M_\phi) R_y(M_\theta) \boldsymbol e_z$, where $R_m(\alpha)$ is a matrix that rotates along the $m$-axis by $\alpha$ and $M = |\boldsymbol M|$. By using the local unitary transformation $H_\text{BdG} \to U H_\text{BdG} U^\dagger - i \hbar U \partial_t U^\dagger$ with $U = e^{+i \sigma_y M_\theta/2} e^{+i \sigma_z M_\phi/2}$, which undoes the $R_i$ rotations at each point in space and time by taking $\boldsymbol M$ back to $\boldsymbol e_z$, we obtain
\begin{align}
H_\text{BdG} = &\left[ \frac{(\boldsymbol p - e \boldsymbol A)^2}{2m^*} + e \phi - \frac{\alpha_R}{\hbar} (\boldsymbol e_z \times \boldsymbol p)\cdot \boldsymbol\sigma - \mu \right]\tau_z + \nonumber\\
&\Delta [\cos\varphi \tau_x - \sin\varphi \tau_y] + M \sigma_z.
\label{eq:H2}
\end{align}
The $\mathfrak{su}(2)$ valued ``four-potential" for spin $\boldsymbol A \equiv i \hbar U \boldsymbol \nabla U^\dagger/e$, $\phi \equiv -i \hbar U \partial_t U^\dagger/e$, or more explicitly
\begin{align}A_i &= +(\hbar/2e)(-\partial_i M_\phi \sin M_\theta, \partial_i M_\theta, \partial_i M_\phi \cos M_\theta ) \cdot \boldsymbol \sigma, \nonumber \\
\phi &= -(\hbar/2e)(-\partial_t M_\phi \sin M_\theta, \partial_t M_\theta, \partial_t M_\phi \cos M_\theta ) \cdot \boldsymbol \sigma,
\label{eq:A-phi}
\end{align}
with $i \in \{x,y\}$ appears as a result of the position and time dependence of the unitary $U$, which yields the covariant derivative $\partial_\mu \to \partial_\mu+ U \partial_\mu U^\dagger$ ($\mu \in \{x,y,t\}$). The term $\boldsymbol A \cdot \boldsymbol p + \boldsymbol p \cdot \boldsymbol A$ that is linear in momentum can be interpreted as a new type of SOC, which is necessary for stabilization of MBS.~\cite{PhysRevLett.105.077001,PhysRevB.81.125318,PhysRevB.82.214509,PhysRevLett.117.077002}

The phase of the pair potential $\varphi$ can be removed by using an additional unitary transformation $V = e^{-i \varphi \tau_z/2}$. If $\varphi$ is inhomogeneous, the derivatives are further modified as $\partial_\mu \to \partial_\mu + i \partial_\mu \varphi \tau_z$, leading to new effective contributions to the exchange field, SOC and single-particle energy. In what follows, we will assume a uniform $\varphi$ such that the overall effect of this transformation is simply $\varphi \to 0$. The resulting Hamiltonian is similar to a well-studied Hamiltonian that is known to stabilize MBS, with a perpendicular magnetic field and an $s$-wave coupling, albeit with an unusual SOC and a renormalized chemical potential.~\cite{PhysRevLett.105.077001,Alicea2011,Alicea2012,Nakosai2013}

We assume the dynamics of the magnetic texture is sufficiently slow to avoid any excitations that can destroy the MBS, such that as the skyrmion in the ferromagnet is moved, the MBS adiabatically follow the magnetic texture. To avoid any excitations via the potential induced by the moving texture, $\phi$, we require that it is negligible compared to $\Delta$, or equivalently $\hbar v_x / R_c^x \ll \Delta$, where $v_x$ and $R_c^x$ respectively denote the speed and typical length of texture variation along the direction of the motion in a racetrack. The adiabatic transport of the MBS further restricts the speed of skyrmions. A rough estimate for this constraint can be obtained using the Landau-Zener formula as $J v_x/R_c^x \ll (E_1-E_0)^2/\hbar$, where $E_0$ and $E_1$ respectively denote the energies of the MBS and the next excited state.~\cite{PhysRevB.90.115404,Landau1932,Zener1932}

\begin{figure}
\centering
\includegraphics[width=\linewidth]{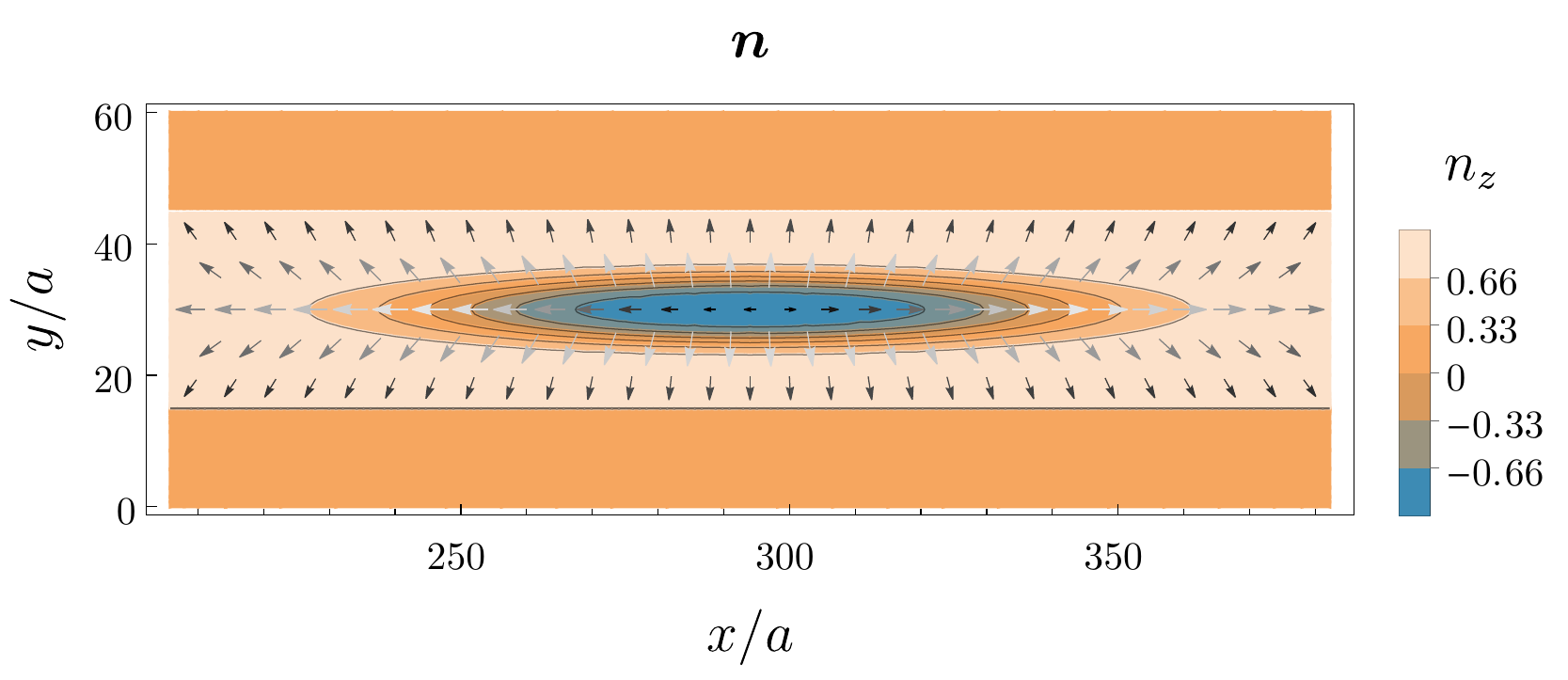}
\caption{The magnetization profile, $\boldsymbol n$, used to represent elongated skyrmions, with horizontal (vertical) core radius $R_c^x/a = 5$ ($R_c^y/a = 49$) and $a=10$nm. Such elongated skyrmions with helicity $\gamma=0$ can be realized in chiral ferromagnets with a combination of Rashba- and Dresselhaus-type DMI. The contours and arrows respectively show the out-of-plane and the in-plane components.}
\label{fig:n}
\end{figure}

The formation of the MBS is determined by the opening and closing of the topological gap. In the presence of spatially varying exchange field and without an extrinsic Rashba SOC, this gap is approximately given by \cite{PhysRevLett.117.077002}
\begin{align}
\frac{E_g}{2} \approx M - \sqrt{ \left( \mu - \frac{\hbar^2 \sum_i \partial_i \boldsymbol M \cdot \partial_i \boldsymbol M}{8 m^* M^2} \right)^2 + \Delta^2 }
\label{eq:topological}
\end{align}
provided that $\boldsymbol M$ is sufficiently smooth. An equivalent expression was previously derived for a rotationally symmetric 2D magnetic texture in Ref.~[\onlinecite{PhysRevB.93.224505}]. The contour $E_g = 0$ determines the topological region \cite{PhysRevLett.105.177002,Kim2015b} which may host MBS depending on the shape of this region.~\cite{PhysRevB.93.224505,PhysRevLett.117.077002}
For example, the geometry of elongated skyrmions in Fig.~\ref{fig:n} makes them suitable candidates for realization of topological regions shaped as wires.
\begin{figure}
\centering
\includegraphics[width=\linewidth]{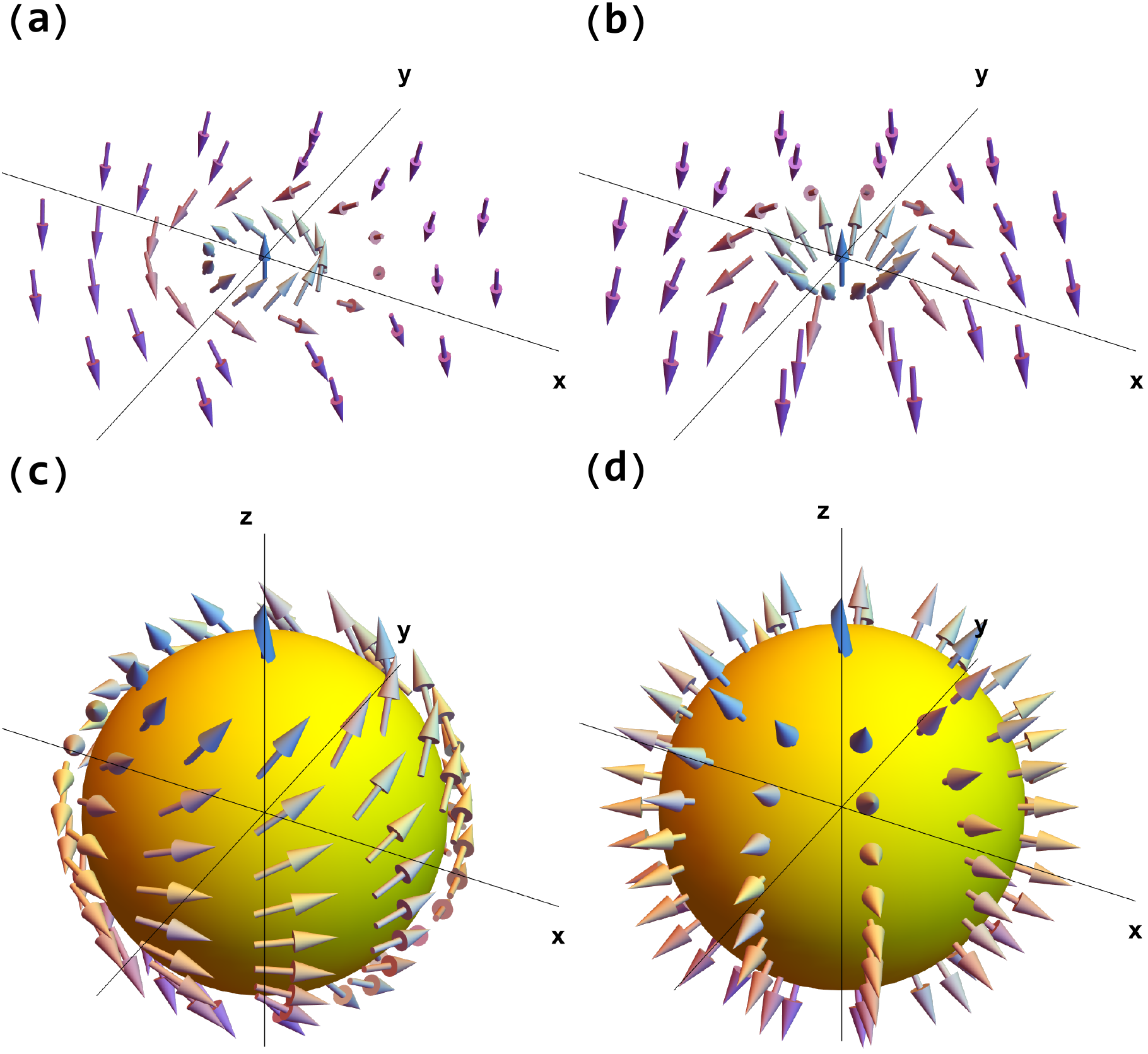}
\caption{Magnetic textures corresponding to Bloch (a) and N\'eel (b) skyrmions with topological charge $Q = 1$. For Bloch (c) and N\'eel (d) skyrmions, the vectors representing magnetization direction wrap around a unit sphere upon application of stereographic projection from two-dimensional space to unit sphere.}
\label{fig:skx}
\end{figure}

\section{Skyrmions stabilizing MBS}
In this section, we describe topological magnetic textures, skyrmions, and show how such textures can be used for stabilizing MBS. We also discuss advantages of skyrmions for MBS manipulation. For more detailed discussion of skyrmions, we refer reader to recent reviews.~\cite{Zhang2020,EverschorSitte2018,Kovalev2018,Jiang2017,Tokura2020,arxiv.2102.10464,Zlotnikov2021} 

\subsection{Description of skyrmions}
A magnetic skyrmion in a quasi-two-dimensional magnetic layer is characterized by an integer invariant, referred to as topological charge. This invariant corresponds to the map from the physical
two-dimensional space to the target space $S^2$ corresponding to magnetization directions.
The topological charge becomes 
\begin{align}
Q = \frac{1}{4 \pi}\int d^2 r (\partial_x \boldsymbol n \times \partial_y \boldsymbol n)\cdot \boldsymbol n.
\end{align}
The topological charge describes how many times magnetic moments wrap around a unit sphere in the mapping in Fig.~\ref{fig:skx}.

Skyrmions can be stabilized in chiral ferromagnets with the Dzyaloshinskii-Moriya interaction (DMI), with the free energy given by $F = \int d^2 r \mathcal F$ where
\begin{align}\label{eq:freeenergy}
\mathcal F = \sum_i \frac{A}{2} (\partial_i \boldsymbol n) \cdot (\partial_i \boldsymbol n) + \boldsymbol D_i \cdot (\partial_i \boldsymbol n \times \boldsymbol n ) - K_u^\text{eff} n_z^2 - \mu_0 M_s H n_z.
\end{align}
Here  DMI is described by a tensor $D_{ij}=(\boldsymbol D_j)_i$, $A$ denotes the strength of the ferromagnetic exchange strength, $K_u^\text{eff} = K_u - \mu_0 M_s^2/2$ is the the effective uniaxial anisotropy with contributions from magnetocrystalline anisotropy, $\mu_0 H$ describes externally applied magnetic field along $z$-axis, and $\boldsymbol n$ is a unit vector along the magnetization direction. DMI originates in the spin-orbit interaction, and its form is determined by the symmetries, including crystal structure and structural symmetries of the multilayer system.~\cite{DzyaloshinskyJPCS1958,MoriyaPR1960} The most common case of Bloch-type skyrmions corresponds to a DMI that is invariant under SO(3) rotations for which $D_{ij} =D_0 \delta_{ij}$, where $\delta_{ij}$ is the Kronecker delta. The case of N\'eel-type skyrmions corresponds to a DMI invariant under SO(2) symmetry which allows only two nonzero tensor coefficients $D_{12}^R=-D_{21}^R$ for two-dimensional magnetic layer, and this case corresponds to Rashba-type DMI. Systems invariant under $D_{2d}$ symmetry can realize Dresselhaus-type DMI for which the nonzero tensor coefficients are related as follows: $D_{12}^D=D_{21}^D$.~\cite{Kovalev2018}  

Depending on the strength of DMI, anisotropy, and strength of magnetic field, chiral ferromagnets can support the formation of skyrmion crystals~\cite{Muhlbauer2009,Yu2010,Koretsune2015,Shibata2015,Gungordu2016a} or formation of isolated skyrmions as metastable states with relatively long lifetime.~\cite{Bessarab2018} Here, we concentrate on isolated skyrmions but topological superconductivity can be also realized with skyrmion lattices~\cite{PhysRevB.91.155405,Mascot2021} and magnetic texture crystals.~\cite{PhysRevB.104.174502,PhysRevResearch.4.013225} The shape of skyrmions is dictated by the form of the DMI tensor. A ferromagnet with anisotropic DMI, which can be represented as a combination of DMI of Rashba- and Dresselhaus-types, naturally supports elongated skyrmions, with the strength of the elongation determined by the relative strength of the Rashba- and Dresselhaus-type DMI.~\cite{Oh2014,Gungordu2016a} An alternative and more flexible method of generating elongated skyrmions is to move circular skyrmions into a narrow racetrack that is thinner than the skyrmion diameter. Skyrmions adapt to the repulsive force exerted by the walls of the racetrack onto the skyrmion by becoming elongated, and the amount of elongation can be controlled by moving skyrmions between different racetracks of varying widths. Such elongated skyrmions have been confirmed using micromagnetic simulations as well as experiments.~\cite{Hsu2016,Jiang2015,Lin2016,Camosi2017,Hoffmann2017}

Skyrmions can be driven using charge currents, surface acoustic
waves, as wells as gradients of strain, magnetic field, or temperature.~\cite{Sampaio2013,Iwasaki2013,PhysRevB.89.064425,Nepal2018,Yokouchi2020,Yanes2019,Zhang2018,PhysRevB.101.064408,Casiraghi2019,Kovalev2012,PhysRevB.89.241101,PhysRevLett.111.067203,Mochizuki2014,PhysRevLett.112.187203} 
Although magnetic domain walls can also be moved using similar methods, one inherent advantage of skyrmions is their flexibility in deforming their shape to adapt to their surroundings, localized nature, and topological protection, allowing them to move even in the presence of pinning sites, disorder, sample edges or other magnetic textures.~\cite{arxiv.2102.10464}

We model the magnetic texture $\boldsymbol n$ of an elongated skymion with topological charge $Q=1$ as stretching of a circular skyrmion ansatz \cite{Altland2010}
\begin{align}\label{eq:anz}
J \boldsymbol n &=  [A(y) J] (\sin n_\theta \cos n_\phi, \sin n_\theta \sin n_\phi, \cos n_\theta ), \\
n_\theta &= 2 \arctan\left(\frac{R_c^2}{(\sigma x)^2 + y^2}\right) + \pi, \qquad n_\phi = \arctan\left(\frac{y}{\sigma x}\right) + \gamma \nonumber
\end{align}
where $\gamma$ is the helicity, $\sigma$ determines the degree of elongation, and $A(y)$ is a step function that modulates the strength of exchange, evaluating to $1$ in the region $W/2 < y < W+W/2$ containing ferromagnet and $0$ outside, see Fig.~\ref{fig:n}. We will refer to $R_c = R_c^y \sim R/2$ as the core radius of the skyrmion (magnetic moments become parallel to the plane with $n_z = 0$ at around $\approx 0.7R_c$).  Circular skyrmions, $\sigma=1$, correspond to the well-known N\'eel-type skyrmions stabilized by Rashba-type DMI, whereas elongated skyrmions with $\sigma \neq 1$ can be stabilized in chiral magnets with a combination of Rashba- and Dresselhaus-type DMI described by $D_{ij}=D_{ij}^R+D_{ij}^D$.~\cite{Gungordu2016a}
Alternatively, skyrmions can be forced into narrow racetracks using field gradients or currents, where they adapt to the pressure from the racetrack boundaries and become elongated.
In our calculations, we smoothen $A(y)$ using a sigmoid function to avoid abrupt increases in $\partial_i \boldsymbol M$ which becomes relevant when computing the $E_g = 0$ contour. The ansatz in Eq.~\eqref{eq:anz} has the advantage of being analytical with well-defined tunable parameters, and it captures all the essential features of elongated skyrmions necessary to analyze the formation of MBS.

We numerically solve the 2D time-independent BdG equation, $H_\text{BdG} \Psi_n = E_n \Psi_n$, for a fixed magnetic texture, neglecting any backaction of the superconductor on the magnet, using Mathematica's NDEigensystem function with Dirichlet boundary condition $\Psi_n(x_b,y_b) = 0$, where $(x_b,y_b) \in \Omega$ and $\Omega$ is the boundary contour. We set the rectangular region containing the 2DEG as $x \in [0,2W/ \sigma]$, $y \in [0,2W]$. In our calculations, we write the Hamiltonian in units of the pairing potential strength $\Delta$, and define the dimensionless tunable parameters $\tilde\mu = \mu/\Delta$, $\tilde B = g^* \mu_B B/2 \Delta$, $\tilde{J} = J/\Delta$, $\tilde {\boldsymbol M} = {\boldsymbol M}/\Delta$ such that
\begin{align}
\frac{H_\text{BdG}}{\Delta} = \left[ -\tilde t (\partial_{\tilde x}^2 + \partial_{\tilde y}^2) -\tilde \mu + i \tilde \alpha_R (\boldsymbol e_z \times \tilde{\boldsymbol \nabla}) \cdot \boldsymbol \sigma \right] \tau_z + \tau_x + \tilde {\boldsymbol M} \cdot \boldsymbol \sigma,
\end{align}
with $\tilde t = \hbar^2/2m^* a^2\Delta$, $\tilde \alpha_R = \alpha_R / \hbar a \Delta $, $\tilde x = x/a$, $\tilde y = y/a$, and we take the effective mass $m^* = 0.1 m_e$, where $m_e$ is the electron mass, $a=10$nm, $W/a=60$ and $\Delta = 0.25$meV, unless stated otherwise.
We first solve this BdG equation using an elongated skyrmion with $R_c^y/a=5$ and $\gamma = 0$ or $\pi/2$, with and without the extrinsic Rashba SOC, and also explore the effects of $\alpha_R$ and the role of helicity $\gamma$. The corresponding magnetic texture for $\gamma=0$ is shown in Fig.~\ref{fig:n}.

\begin{figure}

\begin{overpic}[width=0.85\linewidth]{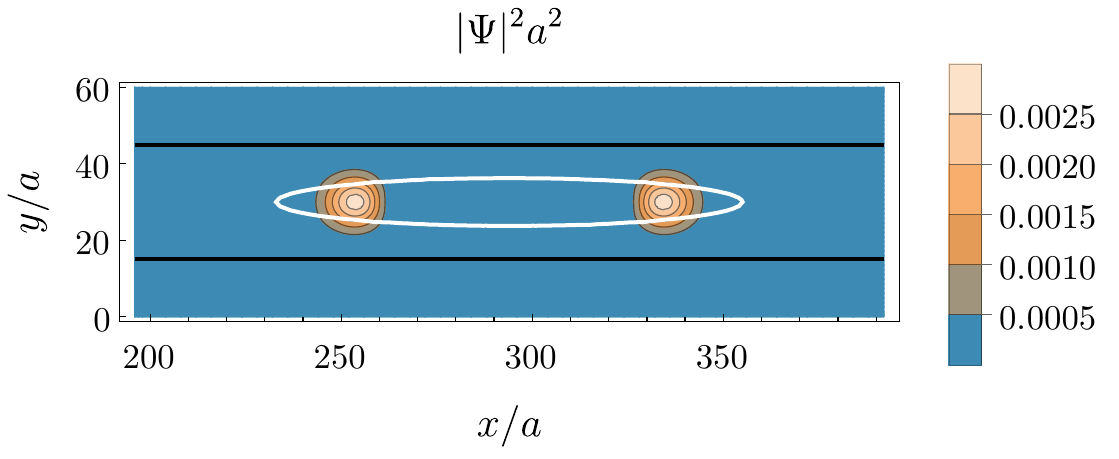}
 \put (-4,40) {(a)}
\end{overpic}

\begin{overpic}[width=0.85\linewidth]{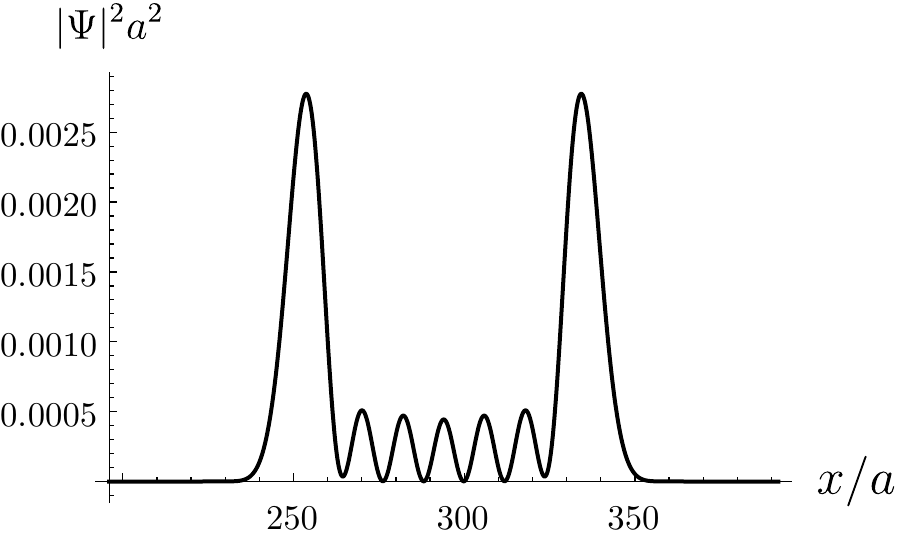}
 \put (-4,62) {(b)}
\end{overpic}

\begin{overpic}[width=0.85\linewidth]{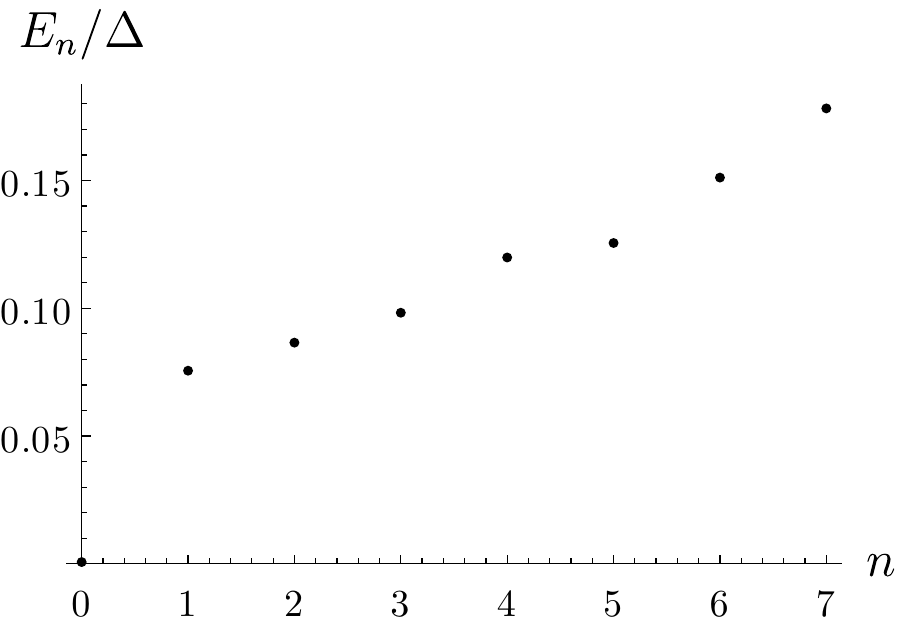}
 \put (-4,72) {(c)}
\end{overpic}

\begin{overpic}[width=0.85\linewidth]{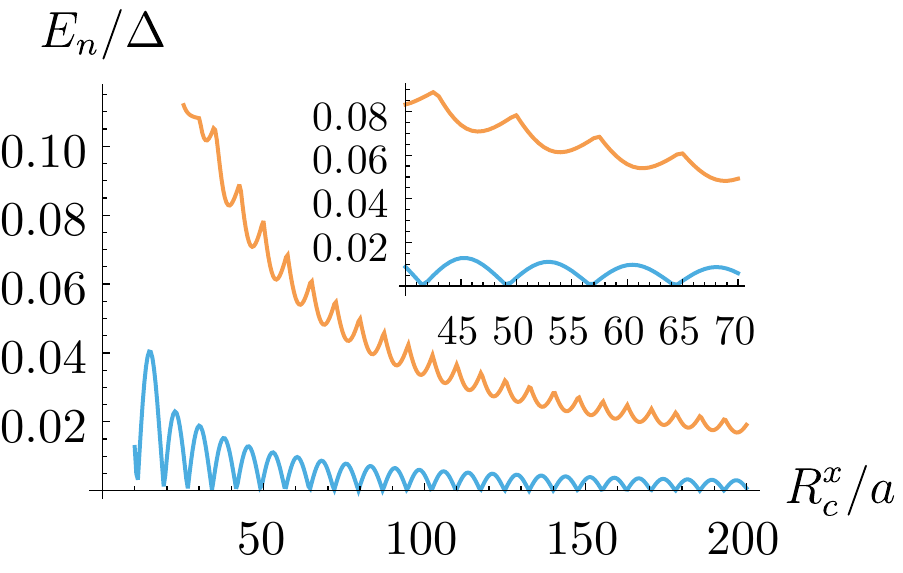}
 \put (-4,64) {(d)}
\end{overpic}

\caption{(a) Probability density $|\Psi|^2$ for the ground state in the $xy$-plane and along a middle cut line $y=W$ in units of $1/a^2$, for a skyrmion with size $R_c^x/a = 49$, $R_c^y/a=5$, using parameters $\tilde B = 0.87$, $\tilde J = 1$, $\tilde\mu=0.2$. The white solid line is the contour enclosing the topological region with $E_g > 0$. The two large peaks localized at the edges of this elliptical region correspond to MBS. (b) Probability density plot along the middle horizontal cut. (c) The low-lying first 8 (nonnegative) energy eigenvalues. (d) Energy of the ground state and the first excited state when using a skyrmion of different elongation as determined by $R_c^x$, using the same parameters otherwise.}
\label{fig:mbs-no-alpha}
\end{figure}
\subsection{MBS localized on elongated skyrmions}
We first analyze the case without extrinsic Rashba SOC ($\tilde \alpha_R = 0$) using a skyrmion with $\gamma=0$. In the first three panels of Fig.~\ref{fig:mbs-no-alpha}, the resulting squared magnitude of the ground state wavefunction $\Psi \equiv \Psi_0$ is plotted in the $xy$-plane and along the $y=W$ line, together with the first 8 energy levels, $E_n/\Delta$, for a skyrmion with $R_c^x/a = 49$. The white contour in the $xy$-plane denotes the boundary $E_g = 0$ containing the topological region, which effectively works as a 1D quantum wire. The parameters used $\tilde B = 0.87$, $\tilde J = 1$, $\tilde\mu=0.2$ are tuned to support the formation of the MBS at the edges of this region, such that $E_0 \approx 0$. The topological protection of the MBS is quantified in terms of the energy gap $E_1 - E_0$ which is less than $\Delta$. For chosen parameters, the gap is  $\approx 0.09 \Delta$. The sensitivity of $E_0$ and $E_1$ to $R_c^x$ can be used to tune MBS by changing the skyrmion geometry.

\begin{figure}

\begin{overpic}[width=0.85\linewidth]{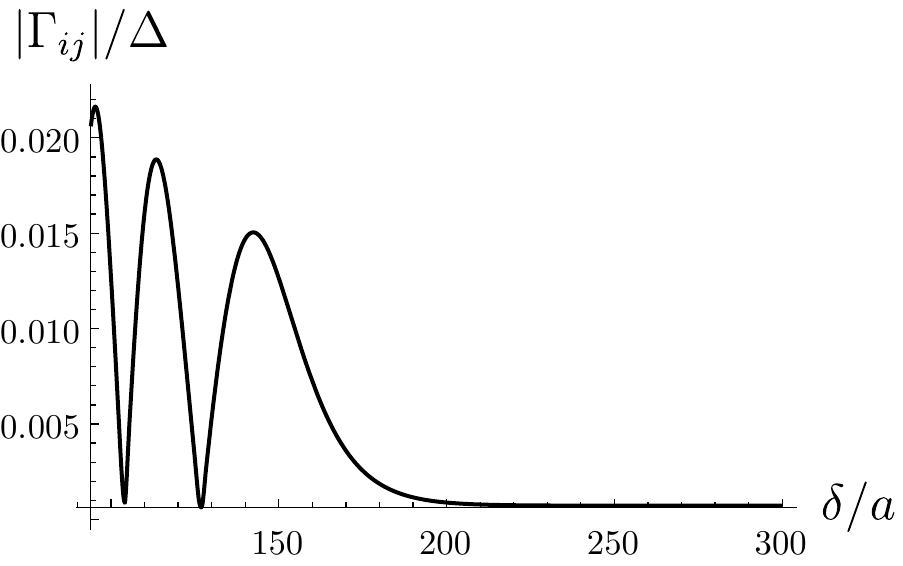}
 \put (-4,63) {(a)}
\end{overpic}

\begin{overpic}[width=0.85\linewidth]{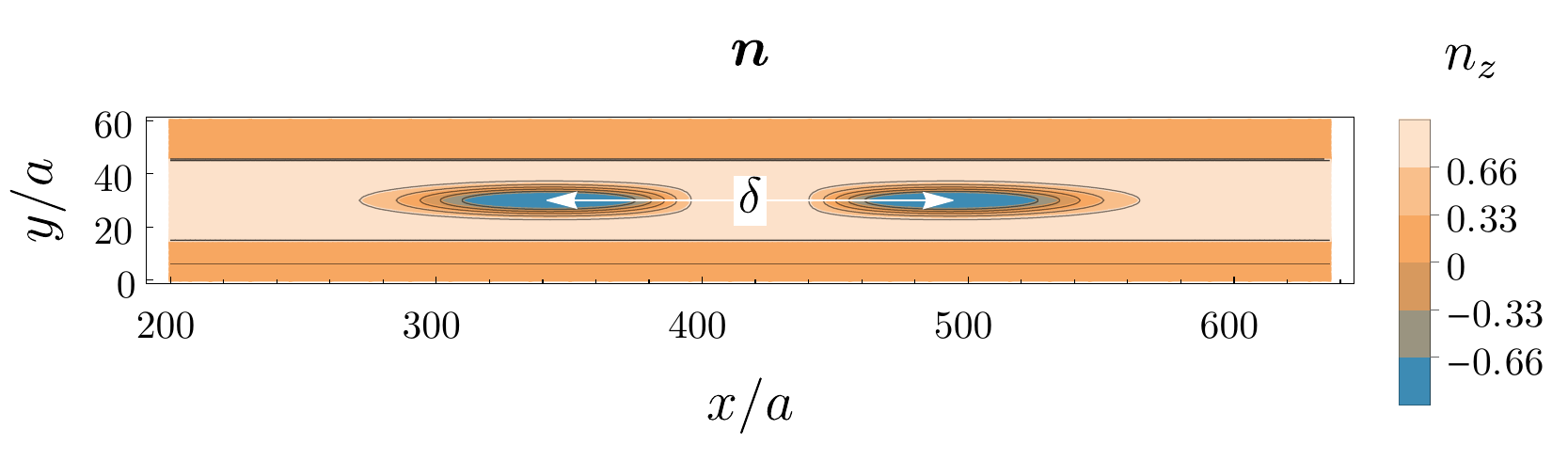}
 \put (-4,25) {(b)}
\end{overpic}

\begin{overpic}[width=0.85\linewidth]{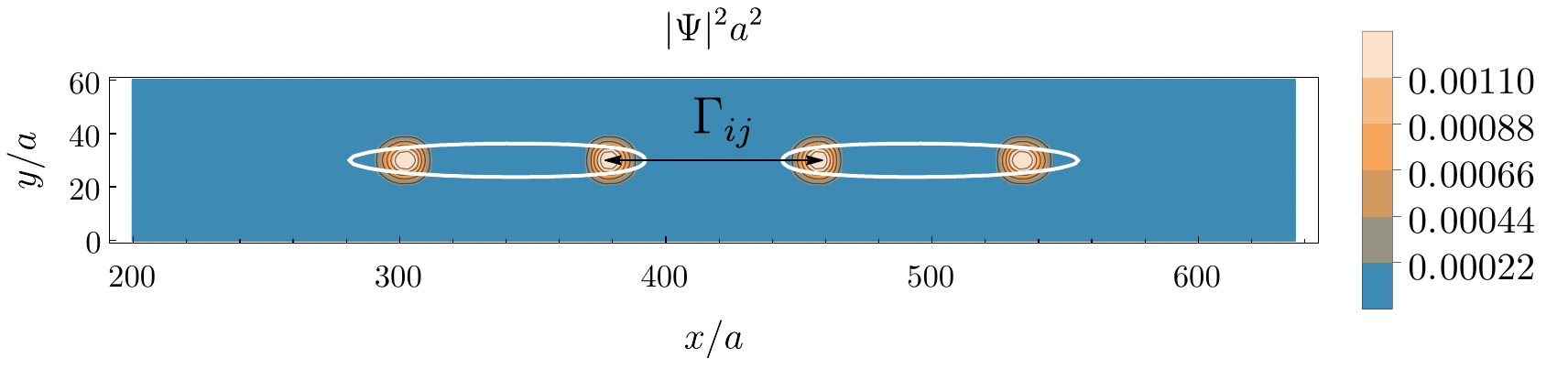}
 \put (-4,21) {(c)}
\end{overpic}

\caption{(a) Hybridization energies $\Gamma_{ij}$ for a pair of MBS hosted at the edges of two distinct elongated skyrmions that are horizontally aligned, as a function of the distance, $\delta$, between the centers of skyrmions along the $x$-axis, in the presence of extrinsic Rashba SOC, obtained from the ground state energy of the 2DEG as the skyrmions are brought closer (using parameters from Fig.~\ref{fig:mbs-no-alpha}). (b) and (c) respectively show $n_z$ and $|\Psi|^2$ for $\delta/a=150$.}
\label{fig:Gamma}
\end{figure}

The last panel of Fig.~\ref{fig:mbs-no-alpha} shows $E_0$ and $E_1$ as a function of the elongation, parametrized by $R_c^x$. Although there are periodic ``good" points at which $E_0 \approx 0$, for a fixed $R_c^y$, we observe that slight perturbations in $R_c^x$ can lead to strong instabilities due to hybridization of MBS for $R_c^x < 10 R_c^y$. This behaviour is due to hybridization of MBS hosted at the edges of the topological region: as the distance between MBS is varied through $R_c^x$, the overlap between MBS oscillates and decays exponentially with the separation. A similar behavior can also be observed between MBS hosted at the edges of different skyrmions, as shown in Fig.~\ref{fig:Gamma}. The same plot also shows a tradeoff: decaying oscillations favor more elongation, but this also leads to decaying of the energy gap, which is detrimental to the robustness of the MBS. The underlying reason for this behavior can be understood from Eqs.~\eqref{eq:H2} and ~\eqref{eq:A-phi}: the effective SOC $\propto \boldsymbol A \cdot \boldsymbol p + \boldsymbol p \cdot \boldsymbol A$ that is required to stabilize MBS is provided by the magnetic texture gradient $\partial_i \boldsymbol M$,~\cite{PhysRevLett.105.077001,PhysRevB.81.125318,PhysRevB.82.214509,PhysRevLett.117.077002} which grows weaker as the skyrmion is stretched further along the $x$-direction: $\boldsymbol A \sim \partial_x \boldsymbol n \sim 1/R_c^x$. We remark that using a skyrmion with $\gamma=\pi/2$ does not affect any of these results. In what follows, we add an extrinsic Rashba SOC to increase the stability of MBS.

\begin{figure}

\begin{overpic}[width=0.85\linewidth]{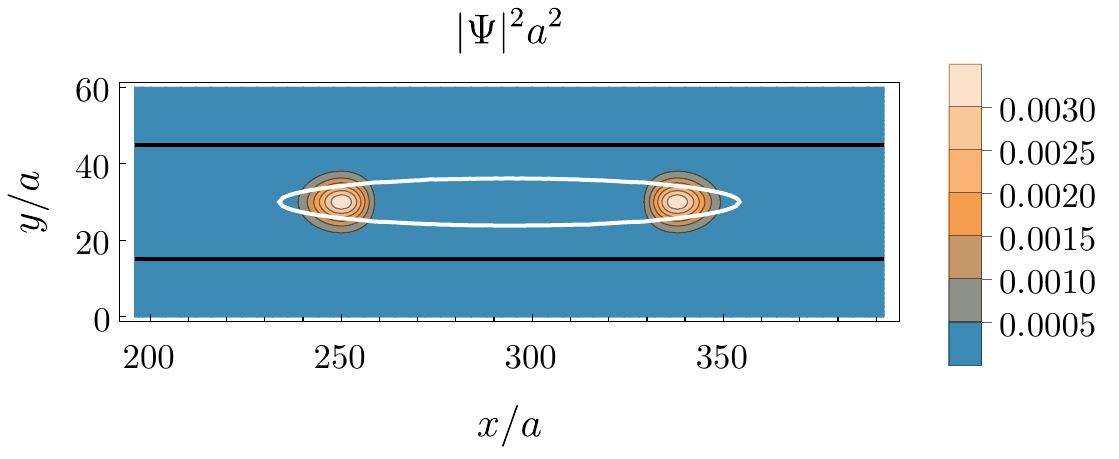}
 \put (-4,40) {(a)}
\end{overpic}

\begin{overpic}[width=0.85\linewidth]{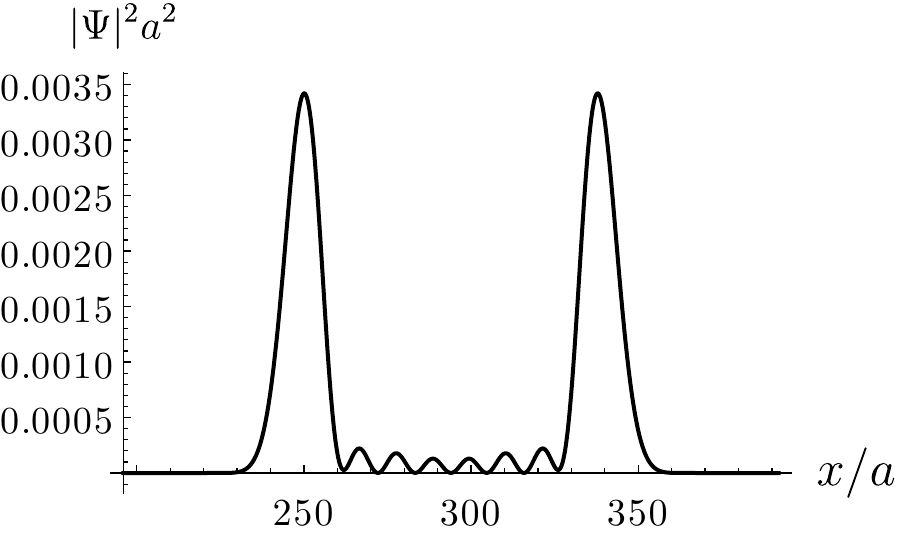}
 \put (-4,62) {(b)}
\end{overpic}

\begin{overpic}[width=0.85\linewidth]{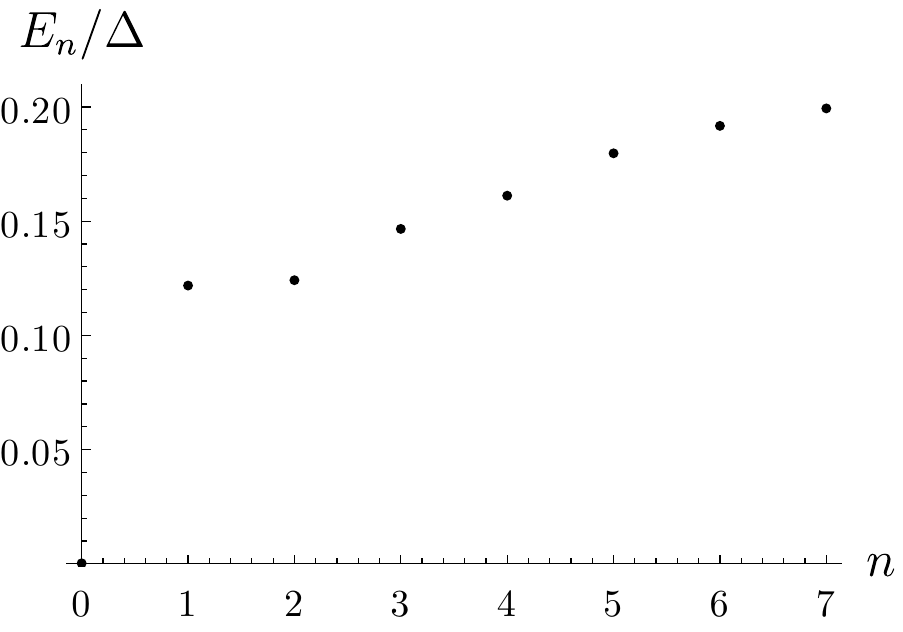}
 \put (-4,72) {(c)}
\end{overpic}

\begin{overpic}[width=0.85\linewidth]{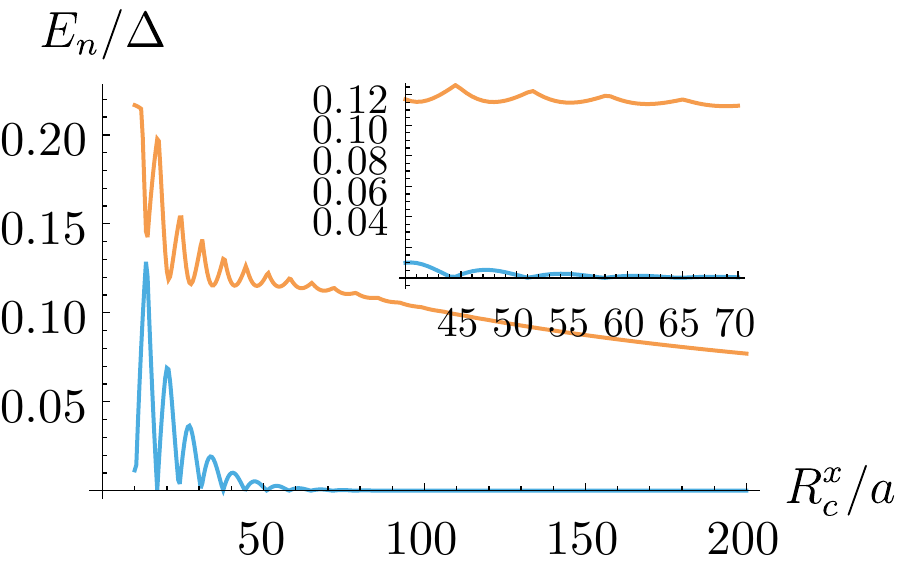}
 \put (-4,64) {(d)}
\end{overpic}

\caption{Similar results as in Fig.~\ref{fig:mbs-no-alpha}, but with extrinsic Rashba SOC $\tilde \alpha_R = 1$.}
\label{fig:mbs-alpha-neel}
\end{figure}

\begin{figure}

\begin{overpic}[width=0.85\linewidth]{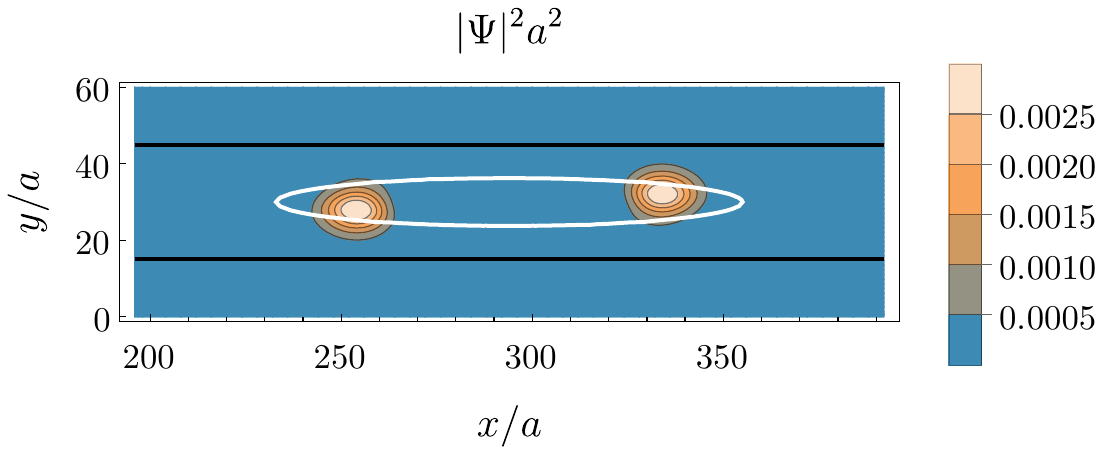}
 \put (-4,40) {(a)}
\end{overpic}

\begin{overpic}[width=0.85\linewidth]{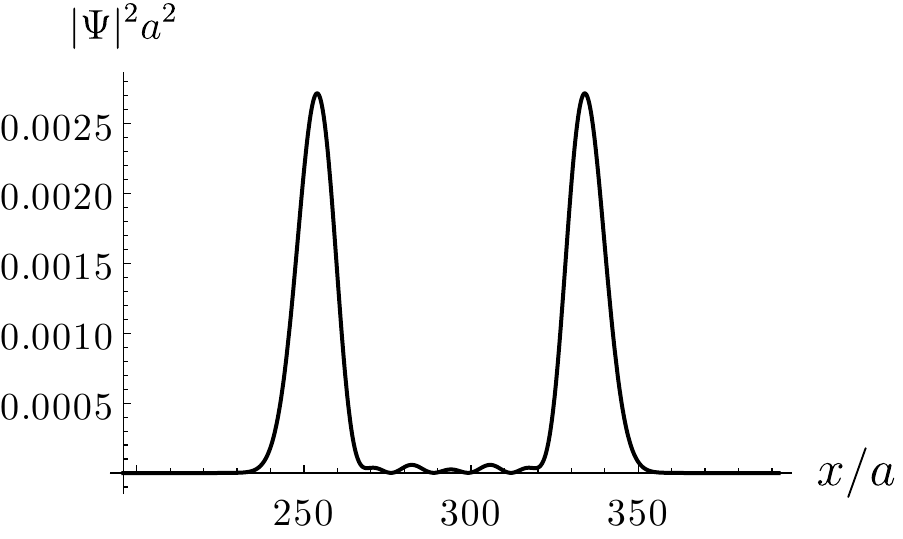}
 \put (-4,62) {(b)}
\end{overpic}

\begin{overpic}[width=0.85\linewidth]{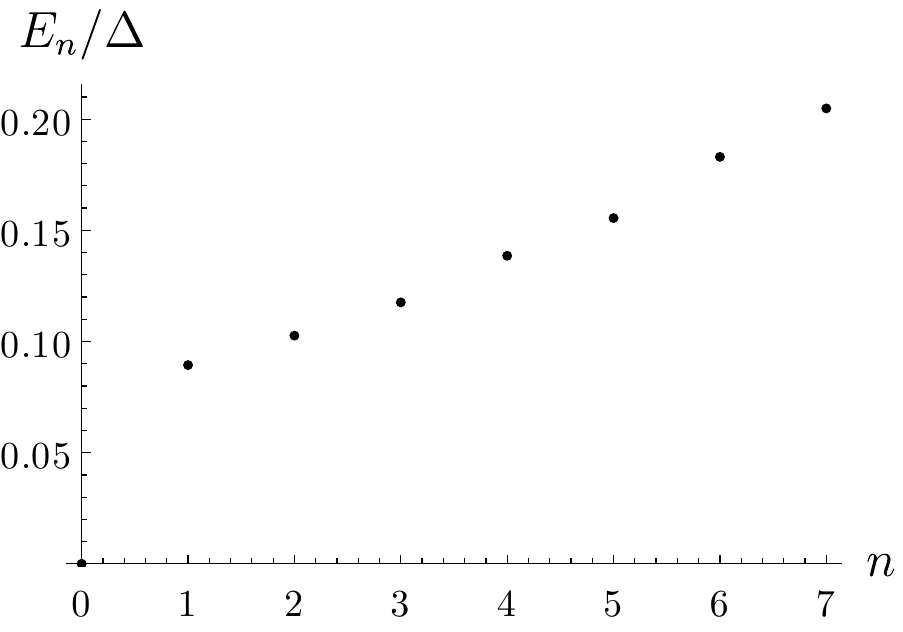}
 \put (-4,72) {(c)}
\end{overpic}

\begin{overpic}[width=0.85\linewidth]{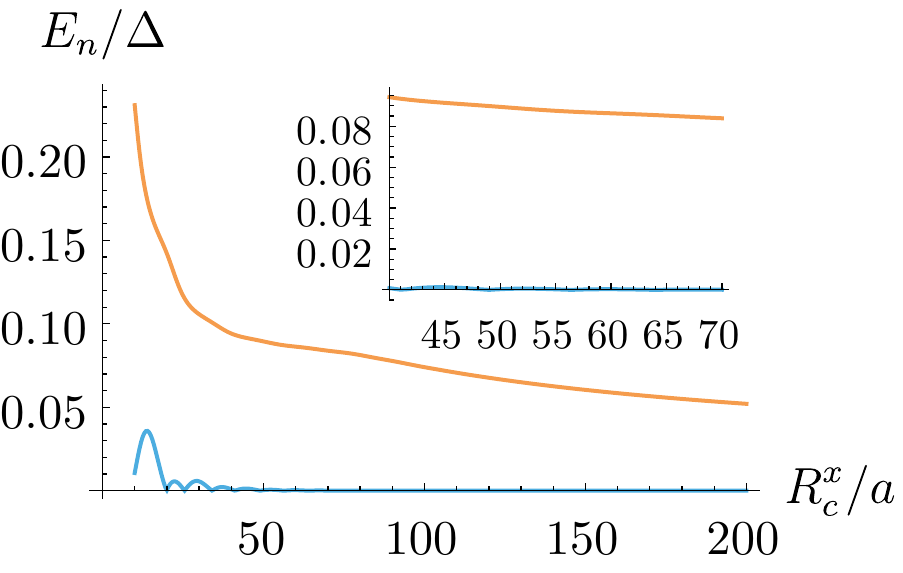}
 \put (-4,64) {(d)}
\end{overpic}

\caption{Similar results as in Fig.~\ref{fig:mbs-no-alpha}, but with extrinsic Rashba SOC $\tilde \alpha_R = 1$ and a skyrmion with helicity $\gamma=\pi/2$.}
\label{fig:mbs-alpha}
\end{figure}

The equivalent results for $\tilde \alpha_R = 1$ (corresponding to $\alpha_R = 2.5$meV nm) are shown in Fig.~\ref{fig:mbs-alpha-neel} for a skyrmion with $\gamma = \pi/2$ (with $\tilde\mu$ changed to 0.28), and in Fig.~\ref{fig:mbs-alpha} for $\gamma=0$. Both sets of results demonstrate the advantages of having an extrinsic Rashba SOC in the 2DEG: the localization of MBS is improved and the oscillations in probability density between the MBS peaks are suppressed compared to the case with $\tilde\alpha_R=0$. Furthermore, the ground state energy remains stable and close to zero for a wide range of elongation values $R_c^x/a > 60$, and the topological gap remains sizable ($E_1-E_0 \sim 0.1 \Delta$) for values of $R_c^x/a$ as large as 200.

For tuning the parameters to support the formation of MBS, we used Eq.~\eqref{eq:topological} as a guide. The spatial profile of the topological region is determined by $\tilde {\boldsymbol M}$, thus, for a given $\tilde J$, we tune the strength of the Zeeman field $\tilde B$ and $\tilde \mu$ to ensure the opening and closing of the gap, and vary the amount of elongation $R_c^x$ to ensure MBS are sufficiently separated spatially. We remark, however, this tuning needs to be redone for a different value of $\tilde J$. This is because changing the value of $\tilde J$ changes the size of MBS, which turns out to be an important parameter. The shape of the ellipse-like topological wire must be such that the region hosts only two MBS, one at each end. When the topological region is too narrow, no MBS form or a pair forms and strongly hybridize~\cite{PhysRevB.90.115404} (a behaviour which can also be observed for the MBS hosted at the ends of two elongated skyrmions as the distance between the skyrmions is reduced, as shown in Fig.~\ref{fig:Gamma}), and when the topological region is too wide, multiple MBS form which again hybridize. For a given $\tilde J$, the shape of the topological region can be adjusted by tuning $\tilde \mu$ and $\tilde B$, or alternatively through $R_c^y$ by fabricating a device with tracks of varying width and using the appropriate ones.

Finally, we remark that a scaling relation $H_\text{BdG} \to \lambda H_\text{BdG}$ exist for the Hamiltonian via the replacements $\{x,y\} \to \{x,y\}/\sqrt{\lambda}$, $\alpha_R \to \sqrt \lambda \alpha_R$, $\Delta \to \lambda \Delta$, $M \to \lambda M$, $\mu \to \lambda \mu$, which can be used to fine-tune the parameters in the presence of experimental constraints.~\cite{PhysRevLett.117.077002,Kim2015b,PhysRevB.93.224505,PhysRevLett.117.077002}

\subsection{MBS localized on chain of circular skyrmions}
\begin{figure}
\includegraphics[width=\linewidth]{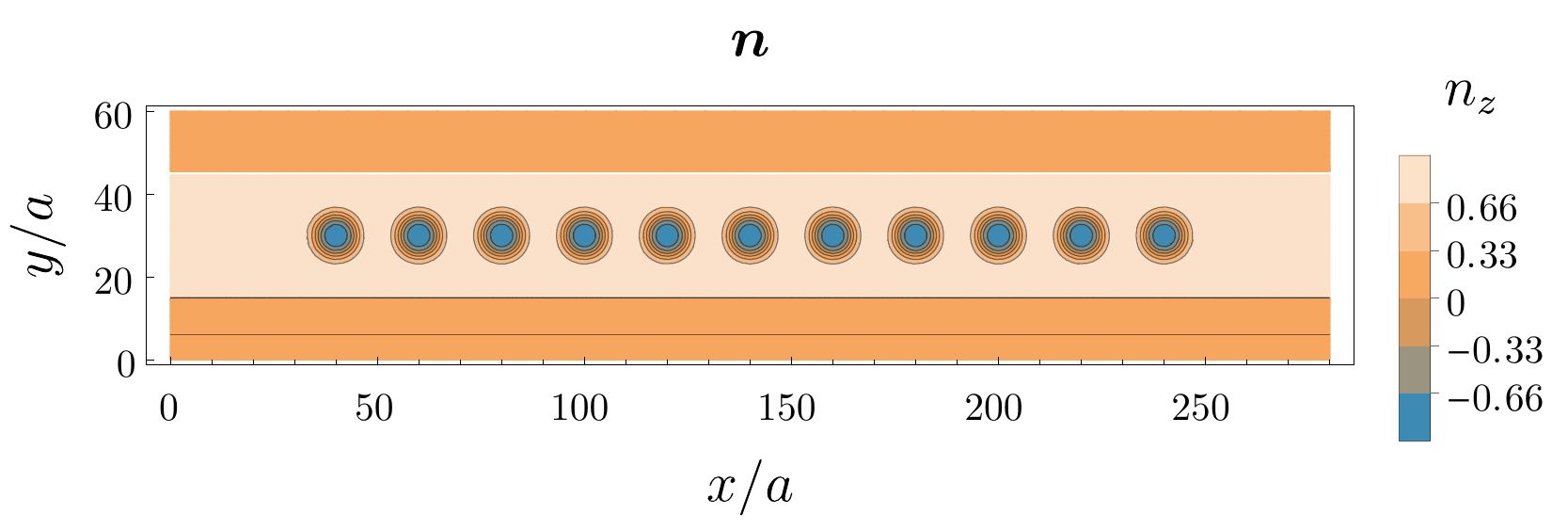}
\caption{Perpendicular ($z$) component of the magnetization profile, $\boldsymbol n$, used to represent a chain of 11 circular skyrmions with core radius $R_c/a = 5$ and spacing $d/a=10$.}
\label{fig:n-chain}
\end{figure}
\begin{figure}
\begin{overpic}[width=0.85\linewidth]{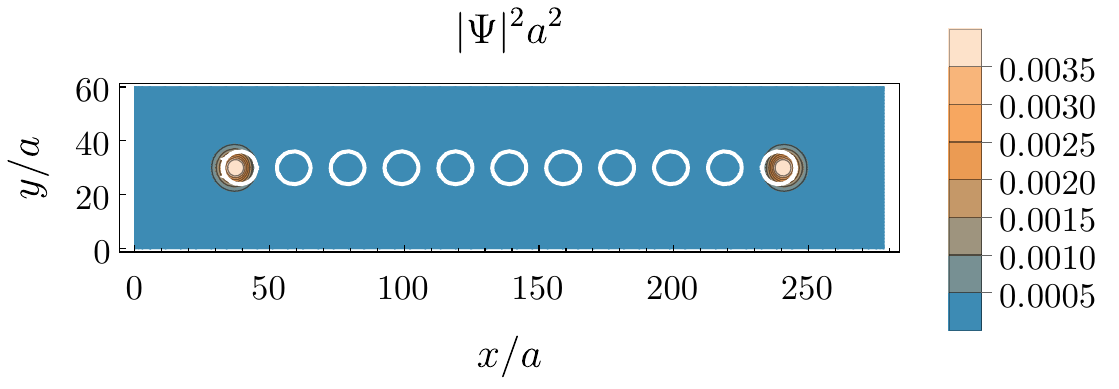}
 \put (-4,34) {(a)}
\end{overpic}

\begin{overpic}[width=0.85\linewidth]{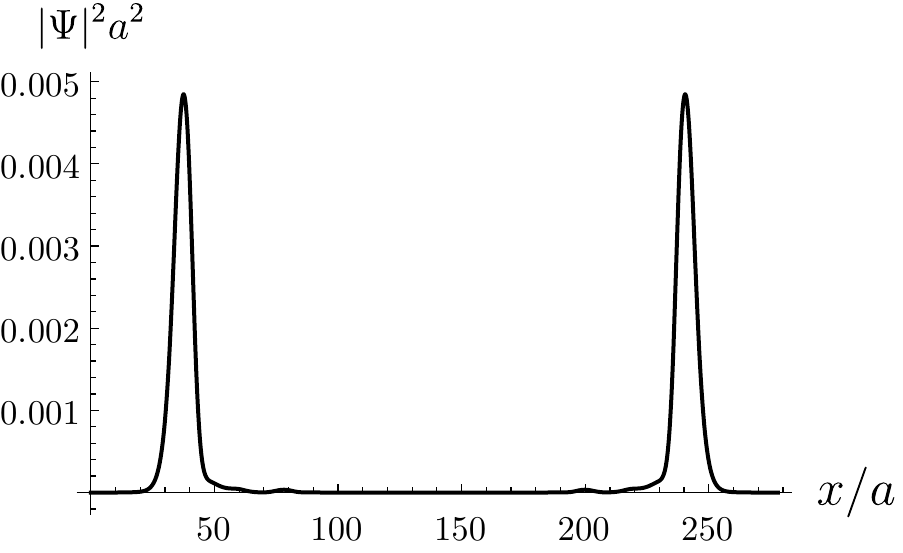}
 \put (-4,62) {(b)}
\end{overpic}

\begin{overpic}[width=0.85\linewidth]{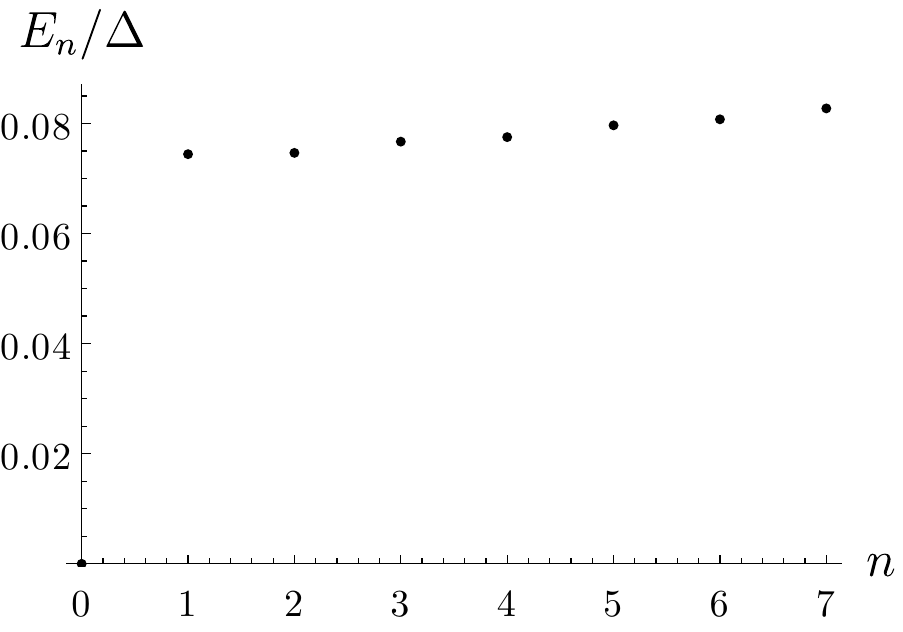}
 \put (-4,72) {(c)}
\end{overpic}

\caption{Similar results as in Fig.~\ref{fig:mbs-no-alpha}, but with $\tilde B=0.8$, extrinsic Rashba SOC $\tilde \alpha_R = 1$, for a chain of 11 circular skyrmions with helicity $\gamma=0$ shown in Fig.~\ref{fig:n-chain}.}
\label{fig:mbs-chain}
\end{figure}

MBS can also be stabilized with a chain of common circular skyrmions with $\sigma=1$. To demonstrate this, we consider a chain of 11 circular skyrmions with core radius $R_c/a = 5$ and helicity $\gamma=0$, interleaved with a spacing of $d/a \approx 13$ (the distance between the centers of two neighboring skyrmions is $\delta=20a$) as shown in Fig.~\ref{fig:n-chain}, and solve the BdG equation in a rectangular region with $W/a=60$ with the extrinsic Rashba SOC using the parameters from the previous subsection, with the exception of $\tilde B = 0.8$. The resulting ground state probability density and low-energy energy spectrum is shown in Fig.~\ref{fig:mbs-chain}. We observe that the contour defined by Eq.~\eqref{eq:topological}, which describes the condition for closing and opening of the topological gap for $\alpha_R = 0$, breaks down in this setup, in which two localized states appear at the ends of the skyrmion chain, similar to the previous results with elongated skyrmions. The resulting zero ground state energy and the gap $\approx 0.08 \Delta$ is robust to perturbations in $R_c$ and $d$, which indicates that this result is not due to an accidental (and unstable) hybridization like in Fig.~\ref{fig:Gamma}. To rule out topologically trivial states with energies accidentally tuned to zero,~\cite{PhysRevB.86.100503,Hess2022} we checked the effect of perturbation in the parameters $\Delta$, $\mu$ and $B$: although the gap slightly varies, the ground state remains zero consistent with MBS, unlike topologically trivial states with exponential sensitivity to such perturbations.~\cite{PhysRevB.86.100503} We obtained qualitatively similar results when we reduced the number of skyrmions to 10.

Due to the repulsive forces between skyrmions, which decay roughly exponentially as a function of the separation distance,~\cite{Lin2013} the skyrmions may tend to spread out. The equilibrium point at which skyrmion dynamics due to repulsive forces stops can correspond to a ratio  $\delta / (2 \times 0.7R_c)\approx 1.25$,~\cite{Zhang2015a} however, it is also possible to use geometric constrictions or applied forces to enforce an inter-skyrmion distance. Another alternative could be to use ferrimagnets or antiferromagnets, which in turn would allow a tighter packing. Similar results were initially demonstrated with antiferromagnetic skyrmions in Ref.~[\onlinecite{PhysRevB.104.214501}] without extrinsic SOC and a gap of $(E_1 - E_0)/\Delta = 0.0016$, where usage of antiferromagnet naturally allows small distances between skyrmions.

\subsection{MBS localized on skyrmions with even vorticity}
\begin{figure}
\centering
\includegraphics[width=0.9\linewidth]{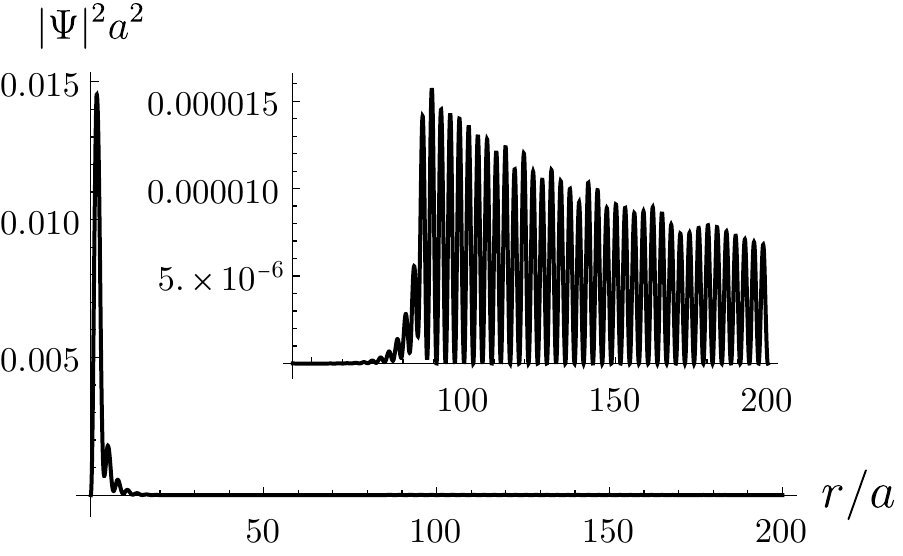}
\includegraphics[width=0.8\linewidth]{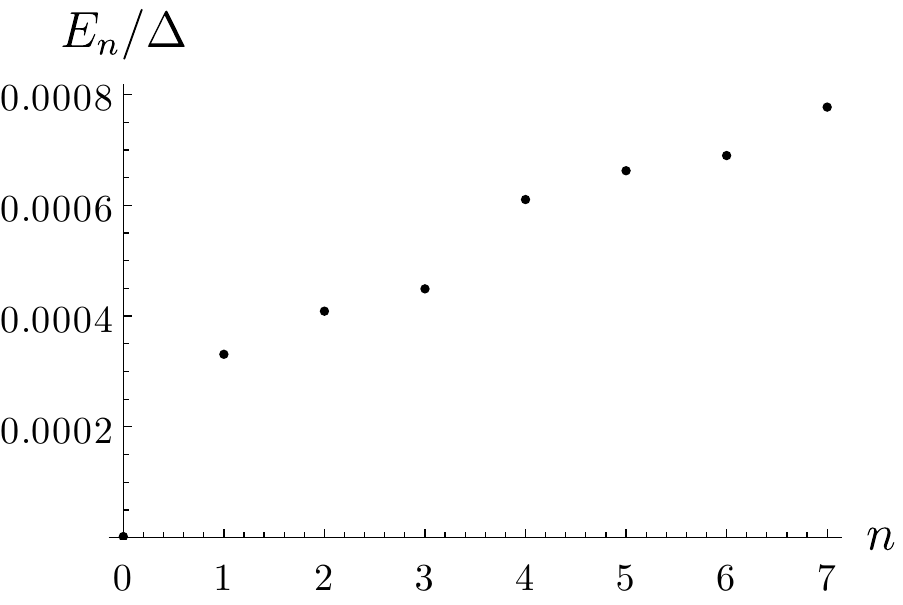}
\caption{Probability density along the radial direction (top) with an inset showing the outer delocalized MBS, and the low energy spectrum (bottom) when using a circular skyrmion  with vorticity $V=2$ and a number of radial magnetization flips $P=25$. The full set of parameters used to solve the BdG equation is given in the text.}
\label{fig:mbs-loss}
\end{figure}

Although a single circular skyrmion commonly realized in chiral magnets with DMI cannot be used to stabilize MBS,~\cite{PhysRevB.93.224505,Pershoguba2016} more exotic circular skyrmions~\cite{Yu2014,PhysRevB.99.064437} with even azimuthal winding number $V$ (vorticity) can support formation of MBS, provided that they also have sufficient number of magnetization flips $P$ along the radial direction to provide hybridization of the modes.~\cite{PhysRevB.93.224505}
\begin{figure}
\centering
\includegraphics[width=0.9\linewidth]{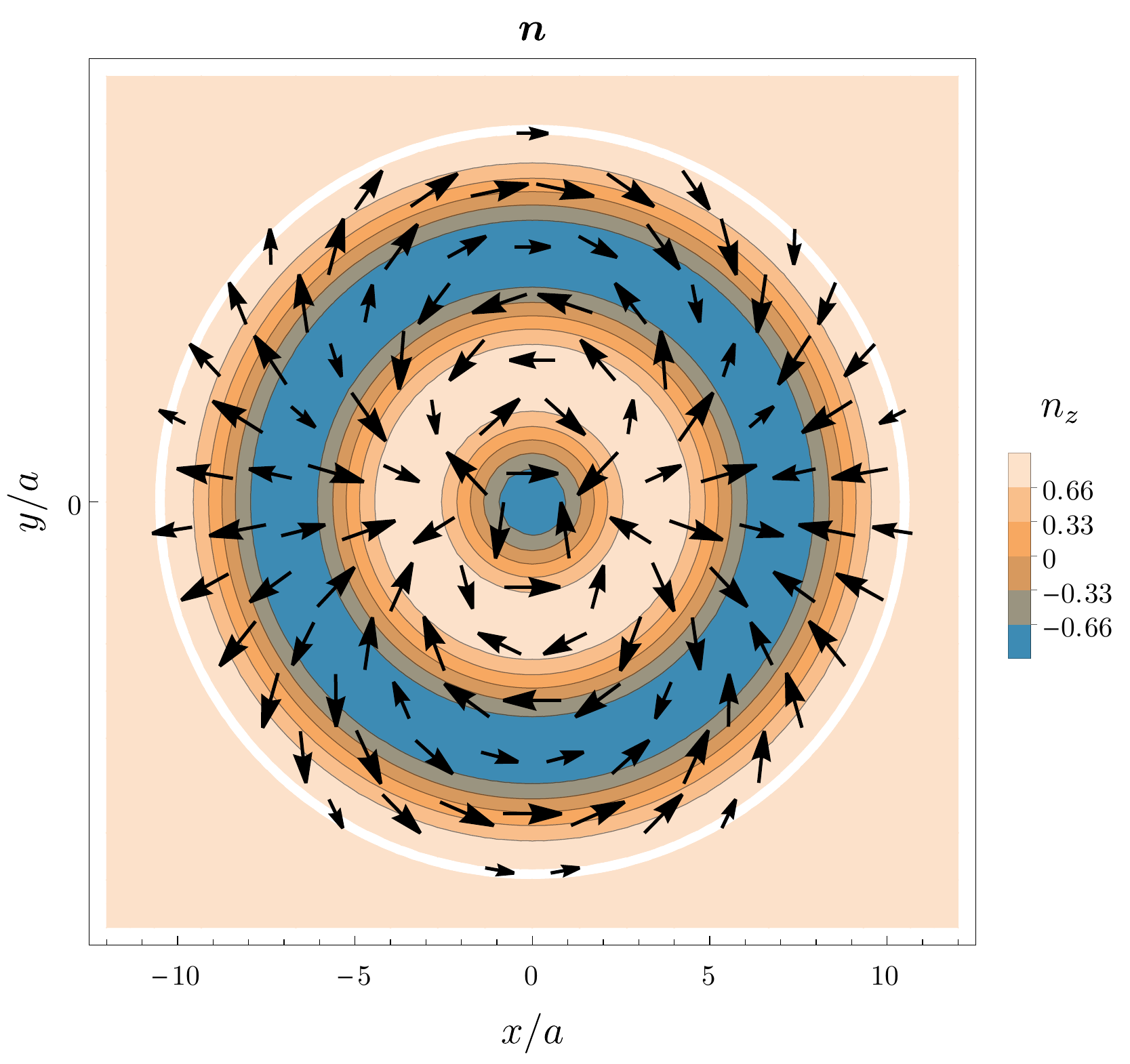}
\caption{Magnetization profile, $\boldsymbol n$, for a skyrmion with even vorticity ($V=2$) and multiple radial magnetization flips ($P=3$), given by  Eq.~\eqref{eq:skyrmion-even-vorticity} with $R_0=0$ and $R/a=3.5$.}
\label{fig:n1}
\end{figure}
We can parametrize such magnetic texture using an ansatz below (see Fig.~\ref{fig:n1}),~\cite{PhysRevB.93.224505}
\begin{align}
n_\phi = V \arctan(x,y), \quad n_\theta =
\begin{cases} 
0 & R_0 < r \\
\pi\frac{r-R_0}{R} & R_0 \leq r \leq R_0 + P R \\
P \pi &r > R_0 + P R
\end{cases}
\label{eq:skyrmion-even-vorticity}
\end{align}
where $r$ is the radial distance from the center of the skyrmion. The topological charge of such skyrmions is given by
\begin{align}
Q = \frac{1}{4\pi} \int d^2 r (\partial_x \boldsymbol n \times \partial_y \boldsymbol n) \cdot \boldsymbol n  = \begin{cases} V & \text{even } P, \\
0 & \text{odd } P.
\end{cases}
\end{align}
Using the parameters: $m^* = m_e$, $\Delta = 0.5$meV, $\mu=0$, $B=0$, $J=1.72$meV, $\alpha_R=0$, $a=5$nm (which approximate the parameters used in Ref.~[\onlinecite{PhysRevB.93.224505}]), we solve the BdG equation in a circular region with radius $200a$, using a skyrmion with parameters: $R/a = 3.5$, $R_0=0$, vorticity $V=2$, and $P=25$.
The resulting probability density and low-energy spectrum are shown in Fig.~\ref{fig:mbs-loss}. Note that the topological gap induced only by the synthetic SOC is relatively small. Combining the synthetic and intrinsic SOC can lead to larger topological gaps as show in Ref.~[\onlinecite{PhysRevB.93.224505}]. We further note that for practical usage of localized MBS one needs to hybridize the outer extended MBS, e.g., by using a pair of skyrmions. We refer to Ref.~[\onlinecite{PhysRevB.93.224505}] for discussion of braiding between the inner modes.

\section{Braiding of Majorana bound states}
In this section, we outline several braiding protocols for MBS localized on skyrmions. A combination of these protocols can allow a realization of initialization, fusion, and braiding of MBS, as required for realizations of MBS signatures relying on non-Abelian statistics.~\cite{Brennen_2009,Alicea2011,Campbell2014,PhysRevB.90.115404,PhysRevB.94.235446,Plugge2017,PhysRevB.95.235305,Zhou2022} A typical realization will then contain an array of MBS where four MBS can form a Majorana qubit as shown in Fig.~\ref{fig:CPB}, and where braiding operations will result in quantum gates applied to Majorana qubits. 

\subsection{Braiding with Cooper pair boxes}
\begin{figure}
\centering
\includegraphics[width=0.9\linewidth]{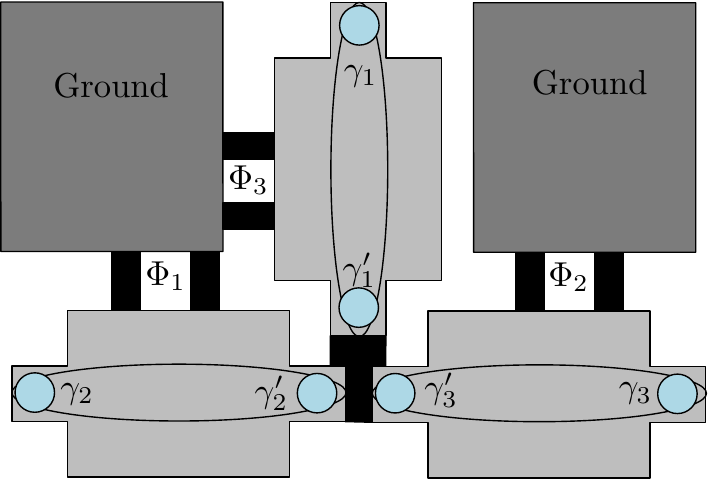}
\caption{Schematic depiction of the superconduction circuit used for the braiding protocol. Cooper pair boxes, each hosting a pair of MBS, are connected to each other with a trijunction and to a bulk superconductor (ground) via split Josephson junctions threading a flux. The Coulomb coupling between the modes on the same island can be controlled by the fluxes.}
\label{fig:CPB}
\end{figure}

Logical operations on topological qubits can be realized through braiding operations, which rely on the non-Abelian statistics of MBS. A typical braiding of MBS is illustrated in Fig.~\ref{fig:braiding}, which implements a nontrivial unitary operation in the low-energy degenerate subspace. Effective braiding can be also
achieved without physically moving MBS by employing tunable couplings between MBS. We adopt this method of braiding to elongated skyrmions by considering a network of superconducting islands acting as capacitors (with capacitance $C$), connected via Josephson junctions (superconductor-insulator-superconductor junctions).~\cite{Sau2011,VanHeck2012} A minimal setup in this proposal involves three elongated skyrmions, each placed on a superconducting island. One side of the islands is connected to the ground (bulk superconductor with very large capacitance) via a pair of split Josephson junctions threading a flux $\Phi_i$, forming three Cooper pair boxes, which are then connected to each other via a trijunction, as shown in Fig.~\ref{fig:CPB}.
We assume that the capacitances are sufficiently large, such that $E_J \gg E_C$ (transmon-like regime~\cite{PhysRevA.76.042319} with reduced charge dispersion) where $E_C = e^2/2C$ is the single electron charging energy of the islands and $E_J = E_J(\Phi_i)$ is the effective Josephson energy of the split junction given by $E_J = 2 E_j \cos(\pi \Phi_i/\Phi_0)$ for symmetric junctions. Here, $E_j$ stands for Josephson energies, $\Phi_i$ is the threading flux, and $\Phi_0 = h/2e$ is the flux quantum. The effective low-energy Hamiltonian for this circuit can be obtained as~\cite{VanHeck2012}
\begin{align}
H_\text{circuit} =& i E_M [\gamma_1' \gamma_2' \cos\alpha_{12} + \gamma_2' \gamma_3' \cos\alpha_{23} + \gamma_3' \gamma_1' \cos\alpha_{31}] \nonumber \\
&- \sum_{j=1}^3 U_j \gamma_j \gamma_j'.
\end{align}
Here, $E_M$ (assumed to be $\ll E_J$) and $U_k \in [U_\text{min}, U_\text{max}]$ are the tunneling and Coulomb energies, and the phase differences $\alpha_{ij}$ are $\alpha_{12} = -(\pi/2\Phi_0)(\Phi_1 + \Phi_2 + 2\Phi_3)$, $\alpha_{23} = (\pi/2\Phi_0)(\Phi_2 + \Phi_3)$, $\alpha_{31} = (\pi/2\Phi_0)(\Phi_1 +  \Phi_3)$. The Coulomb energy $U_k \propto e^{-\sqrt{8 E_J/E_C}}$ has an exponential dependence on flux $\Phi_i$, allowing tunability $U_\text{max} \gg U_\text{min}$ required to turn the interaction on and off. For simplicity, it is also assumed that $E_M \gg U_k$ such that the three modes at the junction fuse into a single one $\gamma_0 \equiv (\gamma_1' + \gamma_2' + \gamma_3')/\sqrt 3$ and we end up with four useful modes $\gamma_1$, $\gamma_2$, $\gamma_3$ and $\gamma_0$, whose interactions can be turned on and off with $U_k$.
The assumption $E_M \gg U_k$ is not essential and simply allows to visualize the braiding procedure better. 

The braiding operation is realized by changing $U_k$, which results in the adiabatic time-evolution of the Heisenberg picture operators $\gamma_i \to \hat U^\dagger \gamma_i \hat U$ where the adiabatic braiding operator is given by~\cite{Sau2011,VanHeck2012}
\begin{align}
\hat U^{(ij)} = \frac{1 + \gamma_i \gamma_j}{\sqrt 2} + \mathcal O \left( \frac{U_\text{min} }{ U_\text{max} } \right).
\label{eq:braiding}
\end{align}
Since $[\hat U^{(ij)}, \hat U^{(jk)}] = \gamma_i \gamma_k \neq 0$, this braiding operation can be used to construct nontrivial quantum gates in an array of such MBS.

The adiabatic sequence for braiding $\gamma_2$ and $\gamma_3$ is given as follows.~\cite{VanHeck2012} Start with $\Phi_1 = \Phi_2 = 0, \Phi_3 = -\Phi_\text{max}$ (where $\Phi_\text{max} < \Phi_0/2$), take $\Phi_1$ to $\Phi_\text{max}$, take $\Phi_3$ to zero. This first sequence transfers $\gamma_2$ to $\gamma_1$. Next, take $\Phi_2$ to $\Phi_\text{max}$, take $\Phi_1$ to zero. This second sequence transfers $\gamma_3$ to $\gamma_2$. Finally, take $\Phi_3$ to $-\Phi_\text{max}$, take $\Phi_2$ to zero, which transfers $\gamma_1$ to $\gamma_3$. During the entire sequence, the modes $\gamma_1$ and $\gamma_1'$ act as ancillaries; at each step, at least one coupling is turned on and one is turned off, and a two-fold energy degeneracy is maintained throughout.

\subsection{Braiding with measurements}
The braiding operation in Eq.~\eqref{eq:braiding} can be also realized by performing projective measurements on MBS.~\cite{PhysRevB.94.235446,Pekker2013,PhysRevLett.116.050501} For the sake of concreteness, we will summarize the scheme introduced in Ref.~[\onlinecite{PhysRevB.94.235446}], in which the topological nanowire (hosting MBS $\gamma_i$ and $\gamma_j$ at its ends) is either coupled to a normal metal loop that threads a flux, or coupled to another topological nanowire via a pair of metal bridges also forming a loop that threads a flux. Transport of particles through the loop happens in conjunction with the bilinear Majorana operator $i \gamma_i \gamma_j$ (which has eigenstates $|\pm\rangle$ with eigenvalue $\pm 1$) with distinct phases. This allows a projective measurement of $i \gamma_i \gamma_j$ either by measuring the conductance of the original topological wire, or by measuring the persistent current through a flux measurement. A braiding protocol can then be constructed using a series of such measurements, which can be described with the following projection operator
\begin{align}
\hat P_{\gamma_i \gamma_j}^{\pm} = \frac{1 \pm i \gamma_i \gamma_j}{2} = |\pm \rangle \langle \pm |.
\end{align}
For example, the following sequence
\begin{align}
P_{\gamma_k' \gamma_k}^{\pm} P_{\gamma_k' \gamma_j}^{\pm} P_{\gamma_i' \gamma_k}^{\pm}|\psi\rangle = \frac{1}{\sqrt 8}\hat U_{ij} |\psi \rangle
\end{align}
leads to the braiding of the modes $\gamma_i$ and $\gamma_j$, provided that all measurements yield $+1$. The sequence is aborted and restarted if any of the measurements is $-1$.
\begin{figure}
\centering
\includegraphics[width=\linewidth]{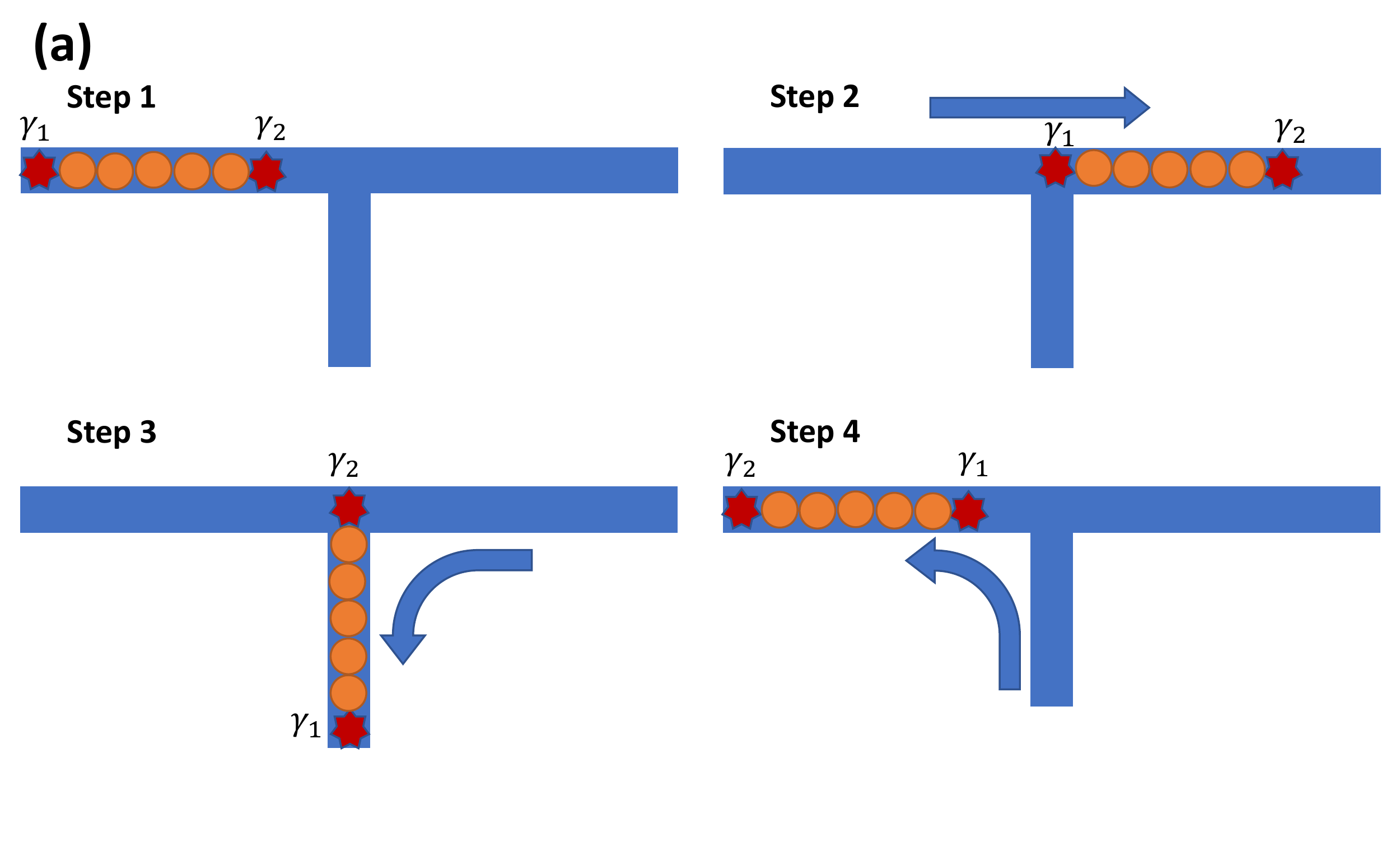}
\includegraphics[width=\linewidth]{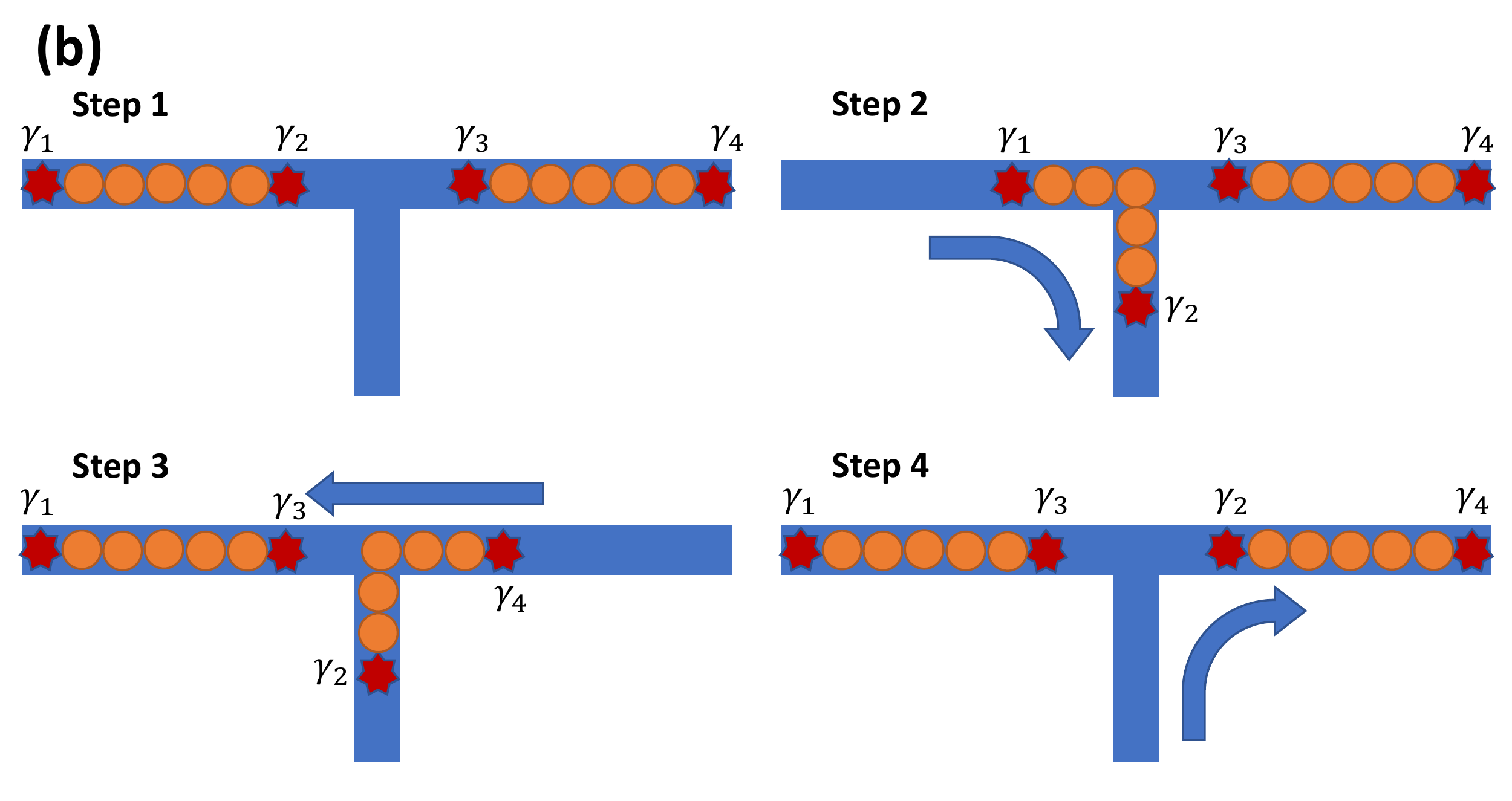}
\caption{An outline of steps realizing adiabatic exchange of MBS across topological (a) or trivial (b) regions using a chain of skyrmions. Arrows indicate directions of adiabatic motion of skyrmion chains with MBS bound to the ends of the chain.}
\label{fig:braiding-sk}
\end{figure}

\subsection{Braiding by moving chains of skyrmions}
Braiding of MBS can be also performed by motion of chains of skyrmions with MBS localized at the ends of chains, see Fig.~\ref{fig:braiding-sk}. Motion of chains of skymions can be driven using charge currents, surface acoustic
waves, as wells as gradients of strain, magnetic field, or temperature.
We can then realize braiding through a topological region, as in Fig.~\ref{fig:braiding-sk}(a), or braiding through a trivial region, as in
Fig.~\ref{fig:braiding-sk}(b). To avoid generation of spurious MBS at the junction, one needs to make sure that the phase difference between different wires meeting at the junction is never equal to $\pi$.~\cite{Alicea2011} From a qualitative analysis, we do not expect that this could happen due to synthetic spin-orbit interaction induced by skyrmions. Numerical calculations performed with the junction in Fig.~\ref{fig:braiding-sk} do not show appearance of spurious zero-modes at the junction. 

In the case of MBS bound to vortices (see Fig.~\ref{fig:vortex}), the MBS operator changes sign when vortex crosses the branch cut. Thus, the exchange of vortices is accompanied by the transformation, $\gamma_1\rightarrow \gamma_2$ and $\gamma_2\rightarrow -\gamma_1$.
MBS in Figs.~\ref{fig:braiding-sk} also undergo such transformation where the Berry phases acquired by the many-body states can be used to explain the origin of this transformation.~\cite{Alicea2011}
A similar braiding procedure was recently proposed in Ref.~[\onlinecite{PhysRevB.104.214501}] for antiferromagnetic skyrmions showing a possibility of tighter skyrmion packing due to usage of antiferromagnet.
Alternatively, one should also be able to use skyrmion realizations in ferrimagnets or synthetic antiferromagnets.~\cite{Legrand2019,Dohi2019}

\section{Conclusions and outlook}
In this tutorial, we have given a pedagogical review of ideas that may lead to realizations of MBS in systems with magnetic textures. We have shown how the Kitaev model of ‘spinless’ fermions can be realized by employing the spin-momentum locking due to spin-orbit interaction. As the synthetic or fictitious spin-orbit interactions can be induced by a texture of magnetic moments, we have emphasized a potential utility
of chiral magnetic textures for realizations of MBS.
In our discussions, we have concentrated on skyrmions as these are well studied magnetic textures that can be easily manipulated using spintronics methods. The most common skyrmions are of Bloch or N\'eel type. Skyrmions can be realized with different shapes, helicities, and vorticities, e.g., antiskyrmions. Skyrmions with charge larger than $1$ have also been predicted~\cite{PhysRevB.99.064437} and can be of use for stabilizing MBS.~\cite{PhysRevB.93.224505,Pershoguba2016} Antiferromagnetic skyrmions have been shown to exhibit certain advantages in stabilizing MBS.~\cite{PhysRevB.104.214501} This shows that further studies of interplay between magnetic textures and proximity effects~\cite{Zutic2019} using synthetic antiferromagnets and ferrimagnets may lead to realizations of MBS in such systems. Two-dimensional magnetic van der Waals materials~\cite{Tong2018,Wu2020,Sierra2021} proximized with other van der Waals materials and superconductors~\cite{Zutic2019} can provide further opportunities in stabilizing MBS.

As experimental signatures of MBS relying on measurements of zero-bias conductance peaks often allow for an alternative explanation involving Andreev bound states or Yu-Shiba-Rusinov states,~\cite{PhysRevB.86.100503,PhysRevB.86.180503,PhysRevLett.109.186802,PhysRevB.96.075161,PhysRevB.96.195430,PhysRevB.98.235406,PhysRevB.97.165302,PhysRevB.98.245407,Avila2019,Prada2020,Kayyalha2020,PhysRevLett.120.156803,Valentini2021} we have also discussed MBS signatures associated with non-Abelian statistics. Since skyrmions can be manipulated using charge currents, surface acoustic waves, as wells as gradients of strain, magnetic field, or temperature, MBS bound to skyrmions can be manipulated in a similar manner. To this end, we have discussed how simple braiding operations can be performed with MBS stabilized by skyrmions. For more complete discussions of MBS braiding protocols, we refer reader to other works~\cite{Alicea2011,PhysRevB.84.035120,PhysRevB.88.035121,PhysRevX.6.031019,PhysRevX.6.031016,PhysRevB.85.144501,PhysRevLett.101.010501,Bonderson2009,PhysRevLett.111.116402,PhysRevB.94.235446,Pekker2013,PhysRevLett.116.050501} and reviews.~\cite{Alicea2012,Beenakker2013,RevModPhys.87.137,Leijnse2012,Sarma2015}

\begin{acknowledgments}
This work was supported by the U.S. Department of Energy, Office of Science, Basic Energy Sciences, under Award No. DE-SC0021019.
\end{acknowledgments}

\section*{Data availability}
The data that support the findings of this study are available from the corresponding author upon reasonable request.

\bibliography{aipsamp,mendeley,extra,skyrmion}

\end{document}